\documentclass[table,xcdraw]{article}

\usepackage[final]{acl}

\usepackage{times}
\usepackage{latexsym}

\usepackage[T1]{fontenc}

\usepackage[utf8]{inputenc}

\usepackage{microtype}

\usepackage{inconsolata}

\usepackage{graphicx}

\usepackage{amsmath}
\usepackage{times}  
\usepackage{helvet}  
\usepackage{courier}  
\usepackage{natbib}  
\usepackage{caption} 
\usepackage{algorithm}
\usepackage{algorithmic}
\usepackage{newfloat}
\usepackage{listings}
\usepackage{booktabs}
\usepackage{multirow}
\usepackage{makecell}
\usepackage{amssymb}
\usepackage{adjustbox}
\usepackage{tabularx}
\usepackage{fancybox}

\usepackage{subcaption}

\newcolumntype{Y}{>{\centering\arraybackslash}X}
\title{Beyond Individual Mimicry: Constructing Human-Like Social network with Graph-Augmented LLM Agents}

\author{
Haoran Bu$^{1}$\thanks{These authors contributed equally.}, Litian Zhang$^{1}$, Chuxuan Zhang$^{1}$\footnotemark[1], Zhanyuan Liu$^{1}$, Hui Pang$^{2}$, Xi Zhang$^{1}$ \\
$^{1}$Cyberspace Security, Beijing University of Posts and Telecommunications \\
$^{2}$Institute of Software, Chinese Academy of Sciences \\
\texttt{\{buhaoran2002, zhanglitian, 2023211472, doubao\}@bupt.edu.cn}, \\
\texttt{panghui@iscas.ac.cn},
\texttt{zhangx@bupt.edu.cn}
}

\begin{document}
\maketitle
\begin{abstract}
Driven by large language models (LLMs), social bot can autonomously engage in local interactions, whose human-like behaviors enable them to evade social bot detection.
However, while these botnets exhibit realistic local social interactions, they fail to preserve human-like social network.
This is because LLM-based bots are graph-unaware and cannot coordinate over global interactions, which makes those botnets vulnerable to graph neural network (GNN)-based detection.
To address this limitation, we propose GraphMind, which equips LLM-driven social bots to explicitly learn and fit human-like social network structures. Building on this foundation, we further construct GraphMind-Botnet, a LLM-driven botnet designed to evaluate the performance of existing social bot detection algorithms. 
Experiments on datasets derived from GraphMind-Botnet show that both text-based and graph-based detection models show substantially degraded performance in distinguishing. Our results highlight the critical role of social link construction in LLM-driven social network generation, while exposing fundamental weaknesses in existing bot detection mechanisms.
\end{abstract}

\section{Introduction}
Recent advances in large language models (LLMs) have fundamentally transformed social bot research. By generating highly fluent, context-aware, and emotionally expressive content, LLM-driven social bots can closely mimic human communication behaviors, substantially narrowing the gap between automated agents and genuine users~\cite{qiao2025botsim,kong2025enhancing}.

Existing LLM-based social botnet simulations primarily focus on enhancing the human-likeness of individual bot behaviors. 
They improves single-agent realism through prompt engineering~\cite{ekin2023prompt} or parameter-efficient fine-tuning techniques~\cite{devalal2018lora,wu2025llm}, enabling bots to produce fluent, emotionally expressive responses and human-like decision-making patterns.

\begin{figure}[t]
  \includegraphics[width=\columnwidth]{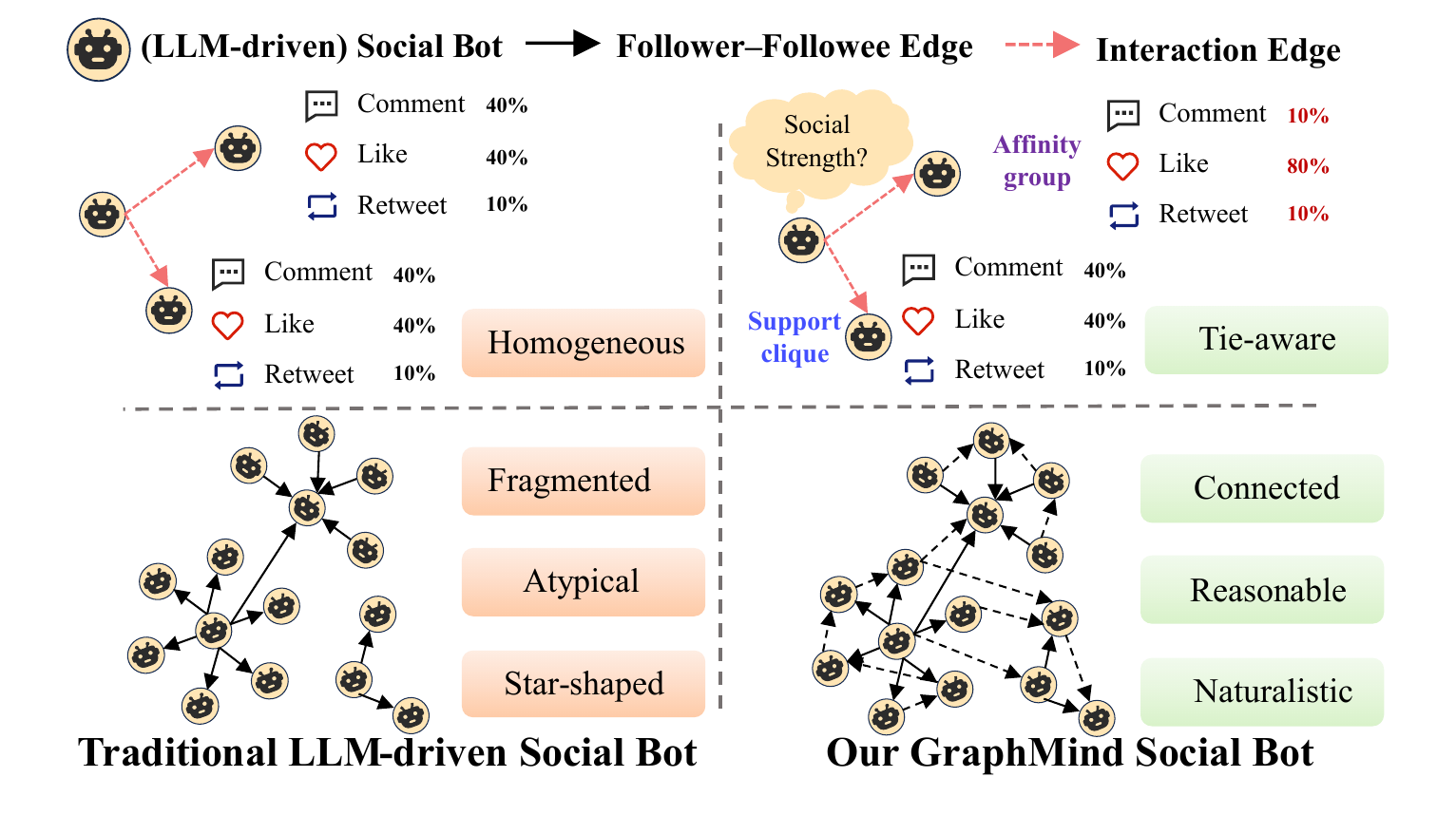}
   \caption{
   The comparation between existing LLM-driven social bot and GraphMind (ours). 
   Our botnet resembles more human-like social networks, leading to significantly improved evasion performance against GNN-based detection.}
   \vspace{-2em}
  \label{fig:method-compared}
\end{figure}

However, while these approaches are effective at the level of individual behaviors and local interaction patterns, they often overlook a fundamental characteristic of real-world social systems: the structure of the underlying social graph.
In existing LLM-driven botnets, individual agents can already be misclassified as human users by detectors that rely primarily on user metadata or textual features~\cite{qiao2025botsim}. Nevertheless, at the network level, such botnets largely fail to establish meaningful and persistent social edges between nodes, which in practice results in a large number of isolated or weakly connected users, and therefore remain vulnerable to graph neural network (GNN)-based detection methods that exploit structural inconsistencies in the social graph~\cite{he2024botdgt,wu2024botscl}.

This vulnerability arises from two key structural deficiencies.
(1) \textit{Existing LLM-driven botnets exhibit superficial behavioral diversity but excessive homogeneity in one-hop interactions.}
In real OSNs, users modulate their interactions based on the strength and context of social ties, leading to heterogeneous engagement patterns across neighbors~\cite{arnaboldi2016ego}.
In contrast, although LLM-driven social bots support a richer set of interaction actions, they tend to apply these actions in a largely uniform manner across neighbors.
Such uniformity induces abnormal ego-network patterns that are readily captured by heterogeneous GNN-based detection methods.
(2) \textit{Current LLM-driven botnets largely lack meaningful multi-hop follow relationships.}
Most LLM-driven bots operate within a small number of local communities, resulting in fragmented graphs with many isolated nodes and weak cross-community connectivity.
This sharply contrasts with real social networks, which are known to be almost fully connected: Facebook reports that 99.91\% of users belong to a single giant connected component~\cite{ugander2011anatomy}.

These limitations arise from a fundamental mismatch between LLMs and social networks. 
LLMs are designed for sequential text modeling and are typically optimized without explicit objectives or state representations over global network topology~\cite{bei2025graphs,tabassum2018social}.
As a result, while individual interactions may appear realistic, LLM-driven botnets fail to jointly optimize graph-level properties, leading to structurally incomplete social networks.

\begin{figure*}[t]
    \centering
    \includegraphics[width=\textwidth]{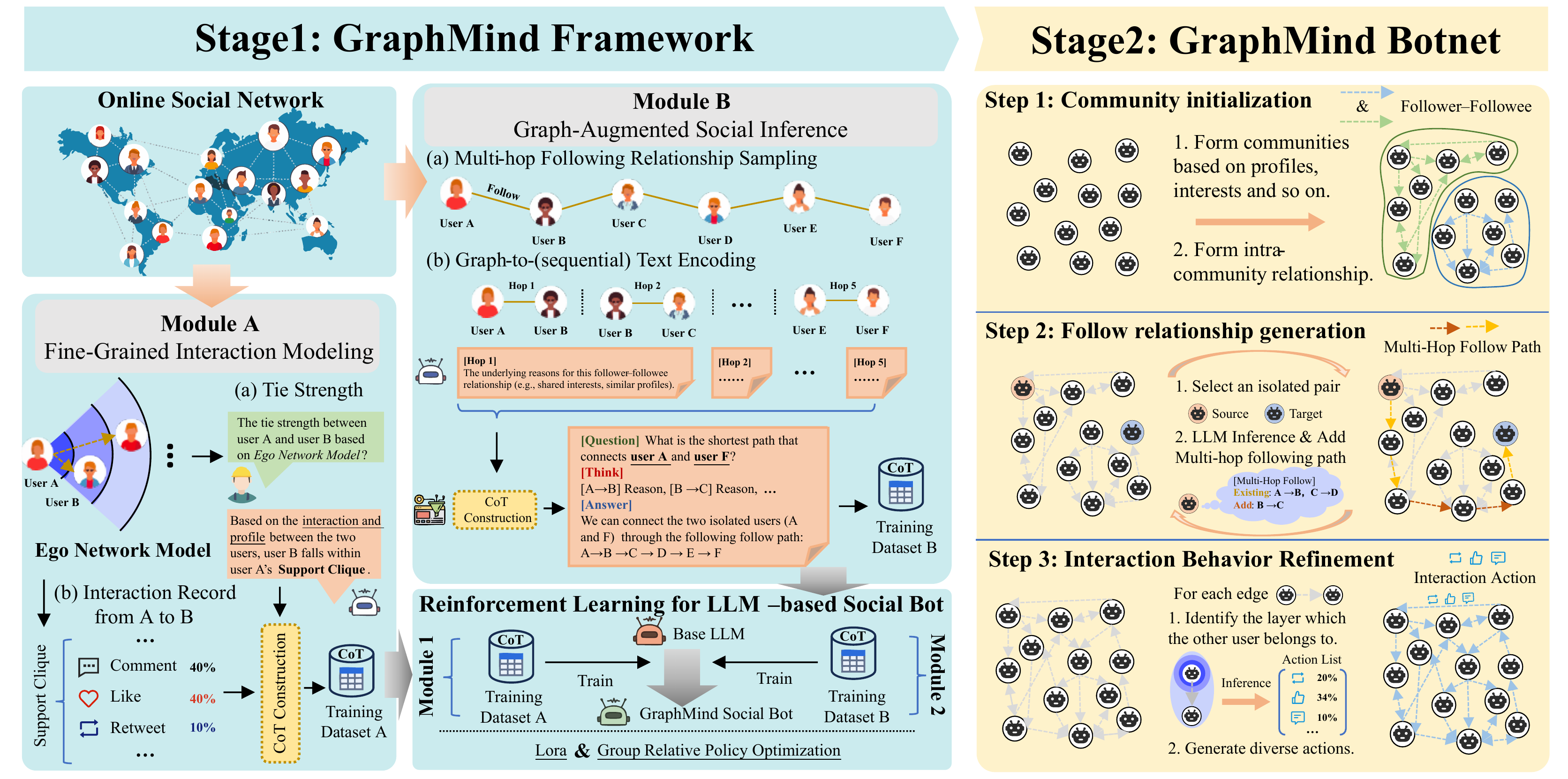}
    \caption{Overview of the framework.
    \textbf{(Left)} GraphMind Framework, where two modules are employed to enable LLMs to generate diverse, strength-aware interactions and to construct multi-hop follow chains, thereby mitigating isolated nodes.
    \textbf{(Right)} Botnet simulation, in which GraphMind social bots autonomously build human-like social networks, improving structural realism and enhancing robustness against GNN-based detection methods.}
    \label{fig:overall_frame}
\end{figure*}

To bridge this gap, we introduce explicit social graph knowledge into LLM-driven social bots by encoding graph information into structured, sequence-based representations compatible with LLM reasoning. We propose \textbf{GraphMind}, the first framework that integrates graph reasoning into LLM-based social behavior modeling. As shown in Figure~\ref{fig:overall_frame} (left), GraphMind operates through two modules: (1) \textit{Fine-Grained Interaction Modeling}, which serializes one-hop relational context (e.g., tie strength) to enable relationship-aware actions, and (2) \textit{Graph-Augmented Social Inference}, which encodes multi-hop structural information into relational sequences, allowing bots to construct cross-community links guided by small-world principles~\cite{milgram1967small} in online social networks.

These agents collectively form the GraphMind botnet, as illustrated in Figure~\ref{fig:overall_frame} (right).
We generate the botnet through a three-stage process: first, we partition agents into communities and construct an initial intra-community network structure; second, GraphMind agents infer potential social edges (specifically, follow relationships) between previously unconnected user pairs; finally, agents generate interaction records conditioned on varying relationship strengths with their neighbors.
The resulting network exhibits improved structural realism compared to prior LLM-driven botnets, substantially degrading the performance of existing GNN-based detection methods.

Our main contributions are summarized as follows:
\begin{itemize}
    \item \textbf{GraphMind Framework.} 
    We present the first systematic study of how LLM-based agents can learn human social network structures and proactively establish social connections during simulation, enabling the generated networks to closely match key structural properties of real-world social graphs.

    \item \textbf{GraphMind Dataset.} 
    Based on GraphMind, we construct an LLM-based dataset whose bots exhibit more human-like network structures than existing LLM-based social simulation, enabling a controlled yet realistic evaluation of the robustness of social bot detection methods.

    \item \textbf{Evaluation and Defense Insights.} 
    Experiments on our dataset demonstrate substantial performance degradation of state-of-the-art detectors, motivating the development of more robust social bot detection approaches.
\end{itemize}

\section{Preliminaries}
\subsection{Problem Formulation}
We model an online social network as a directed graph $G = (V, E)$, where $V$ denotes a set of social bots and $E \subseteq V \times V$ represents directed follow relationships.
Each node $v \in V$ is associated with a textual profile, encoded into a fixed-dimensional embedding $\mathbf{x}_v \in \mathbb{R}^d$ using the embedding layer of an LLM with last-token pooling.
We denote the collection of node representations as $X = \{\mathbf{x}_v \mid v \in V\}$.

Beyond structural links, follow edges may carry interaction annotations such as likes, retweets, and comments, denoted as
\begin{equation}
Y = E_{\text{Interaction}},
\end{equation}
which together with $E$ yields a heterogeneous representation of social relationships.

We assume an initial graph $G_0 = (V, E_0, Y_0)$, where $E_0$ is sparse and $Y_0$ is incomplete, resulting in fragmented structure and limited edge-level semantics.

\subsection{Social Graph Completion}

Given the initial graph $G_0 = (V, E_0, Y_0)$, our objective is to infer missing follow edges
$\Delta E \subseteq (V \times V) \setminus E_0$
\emph{and} their associated interaction annotations $\Delta Y$, such that the completed graph
\begin{equation}
G' = (V, E_0 \cup \Delta E, Y_0 \cup \Delta Y)
\end{equation}
resembles real-world human social networks in terms of both structural and interaction-level properties.

Specifically, we aim to jointly generate $\Delta E$ and $\Delta Y$ so that key graph statistics of $G'$, including connectivity, degree heterogeneity, and short average path length, are close to those observed in real online social networks, while maintaining realistic interaction patterns along inferred social ties.

\section{Methodology}
As illustrated in Figure~\ref{fig:overall_frame} (left), GraphMind comprises two complementary modules that operate at different granularities of social behavior.
Fine-Grained Interaction Modeling generates relationship-aware interaction records conditioned on social tie strength, completing missing edge-level interactions $\Delta Y$.
Graph-Augmented Social Inference enables LLM-driven agents to infer missing follow relationships during simulation, completing the social graph structure by generating $\Delta E$ and jointly yielding networks that are realistic in both structure and interaction dynamics.

GraphMind framework is trained using Lora fine-tuning~\cite{hu2022lora} and {Group Relative Policy Optimization} (GRPO)~\cite{shao2024deepseekmath}. In the following, we describe how to construct the dataset and design the reward function to achieve this objective.
\subsection{Fine-Grained Interaction Modeling}
Interaction-level heterogeneity along social ties plays a critical role in shaping realistic ego-network patterns. In existing LLM-based botnets, interaction behaviors tend to be applied uniformly across neighbors, yielding one-hop interaction distributions that are overly simplistic and fail to capture the diversity of human social interactions. This module is designed to endow agents with relationship-aware interaction behaviors, thereby generating diverse and socially coherent edge-level interactions that better reflect human-like interactions.

To this end, we leverage human interaction data from TwiBot-22~\cite{fengtwibot} as supervision. Each training instance consists of a user pair and their observed interactions, formalized as
\begin{equation}
Y_{\text{fine}} = \{ (\mathbf{x}_u, \mathbf{x}_v, e_{\text{interaction}}, l^e) \mid (u \rightarrow v) \in E_{h \rightarrow h} \},
\end{equation}
where $\mathbf{x}_u$ and $\mathbf{x}_v$ denote the textual profile embeddings of the two users, $l^e$ is the ground-truth relationship strength label defined according to the four-level ego-network hierarchy in Table~\ref{tab:rel_circles}~\cite{arnaboldi2016ego}, and $e_{\text{interaction}}$ represents the sequence of observed actions along the directed edge. Based on these samples, we adopt a chain-of-thought reasoning paradigm~\cite{wei2022chain} to factor interaction generation into two stages: the model first infers latent social relationship attributes from profile representations, and then generates interaction records in the MCP format~\cite{2025Model} conditioned on the inferred relationship context. This two-stage design allows the model to disentangle relationship inference from action generation and produces interaction sequences that are both semantically and socially coherent.

We optimize the model via GRPO to jointly learn relationship inference and interaction generation. For each training sample, the model predicts a relationship level $l^e_{\text{pred}}$ from the profile embeddings and generates a corresponding action sequence. Two complementary reward terms guide this process. The relationship inference reward
\begin{equation}
R_{1} = \exp\!\big(-\lvert l^e_{\text{pred}} - l^e \rvert\big)
\end{equation}
favors predictions that preserve relative social distance with respect to the ground-truth label $l^e$, ensuring that inferred relationships remain consistent with the observed ego-network hierarchy. To enforce action–relationship consistency at the instance level, we measure the divergence between the model-generated action distribution and the empirical distribution of the current sample,
\begin{equation}
R_{2} = -\mathrm{KL}\!\left(
P_{\theta}(\cdot \mid l^e_{\text{pred}})
\,\middle\|\,
P_{\text{sample}}(\cdot)
\right),
\end{equation}
where $P_{\theta}(\cdot \mid l^e_{\text{pred}})$ denotes the empirical distribution derived from the model's generated action sequence, and $P_{\text{sample}}(\cdot)$ denotes the empirical distribution from the ground-truth sequence $e_{\text{interaction}}$. By maximizing $R_{2}$, the model is encouraged to generate action sequences whose distribution closely follows that of the observed human interactions, conditioned on the predicted relationship level. The combined reward is
\begin{equation}
R_{\text{Fine}} = R_{1} + R_{2},
\end{equation}
which drives the model to produce socially plausible and relationship-aware interactions while preserving the diversity and heterogeneity observed in real-world user behaviors. An example of the constructed training sample is provided in Table~\ref{tab:fim-template}.

\subsection{Graph-Augmented Social Inference}
Inspired by WalkLM~\cite{tan2023walklm}, we propose Graph-Augmented Social Inference to address a key limitation of existing LLM-based botnets: the lack of global connectivity and meaningful cross-community social ties. The core challenge lies in reconciling the non-sequential nature of graph structures with the sequential input paradigm of LLMs. To this end, GSI adopts a two-stage training strategy: (1) supervised fine-tuning on multi-hop human social chains to inject structural priors of real social networks, and (2) policy optimization to generate multi-hop path.

\subsubsection{Supervised Fine-Tuning}
The objective is to endow the LLM with prior knowledge of how human users form long-range social connections in real-world networks. 
Instead of learning from isolated edges, we construct supervision signals based on multi-hop follow paths sampled from human social graphs and serialize them into sequential reasoning traces suitable for LLM training.

\paragraph{Multi-hop Human Relationship Sampling.}
We extract multi-hop follow paths from a human-only subgraph of TwiBot-22 to serve as structural supervision. Specifically, a human user $v_{h_0} \in V_H$ is selected as the anchor node of a social chain $C$, which is iteratively extended along directed human-to-human follow edges. A sampled chain is defined as
\begin{equation}
C = [v_{h_0}, v_{h_1}, \dots, v_{h_k}],
\end{equation}
where each adjacent pair satisfies $e_{v_{h_i} \rightarrow v_{h_{i+1}}} \in E_{h \rightarrow h}$.

The expansion process terminates when no further outgoing human-follow edges are available or when the chain reaches a maximum length. Consistent with empirical observations of small-world social networks~\cite{milgram1967small}, we restrict the maximum chain length to six hops.

To construct semantically coherent and structurally meaningful paths, node selection is guided by both semantic similarity to the anchor user and node influence. At each step, the next node is selected by
\begin{equation} 
    \begin{aligned} 
        v_{h_{i+1}} &= \arg\max_{v \in \mathcal{C}(v_{h_i})} \Big( \cos(\mathbf{x}_{v_{h_0}}, \mathbf{x}_v) + \tilde{d}^{-}(v) \Big), 
    \end{aligned} 
\end{equation} 
\begin{equation}
\tilde{d}^{-}(v) = \frac{\deg^{-}(v)}{\max_{u \in \mathcal{C}(v_{h_i})} \deg^{-}(u)},
\end{equation} 
where $\mathcal{C}(v_{h_i})$ denotes candidate human nodes reachable from $v_{h_i}$, $\tilde{d}^{-}(v)$ denotes normalized in-degree. This strategy favors paths that remain semantically grounded while naturally traversing structurally influential users, which often act as bridges across communities.

\paragraph{Graph-to-Sequential Text Encoding.}
Since LLMs do not natively operate on graph-structured inputs, each sampled multi-hop chain is transformed into a sequential natural-language representation. Given a chain $C = [v_i, \dots, v_j]$, we decompose it into consecutive follow relations. For each hop $(v_k \rightarrow v_{k+1})$, we generate a brief textual rationale $t_k$ explaining why user $v_k$ follows $v_{k+1}$, conditioned on user profile attributes and node-level structural signals (e.g., relative in-degree).
The hop-level rationales are concatenated to form a chain-of-thought (CoT) sequence as bellow
\begin{equation}
\text{COT}_{v_i \rightarrow v_j}
=
\langle \text{User} \rangle_i
\oplus t_i
\oplus \dots
\oplus t_{j-1}
\oplus
\langle \text{User} \rangle_j,
\end{equation}
where $\oplus$ denotes textual concatenation. This representation enables the LLM to learn how long-range social connections are composed through a sequence of locally plausible follow decisions.

\paragraph{Loss Function.}
Let $\mathcal{D}_{\text{SFT}}$ denote the constructed dataset of multi-hop reasoning sequences. The supervised fine-tuning objective minimizes the negative log-likelihood of the reference CoT sequences:
\begin{equation}
\mathcal{L}_{\text{SFT}}
=
-\frac{1}{|\mathcal{D}_{\text{SFT}}|}
\sum_{C \in \mathcal{D}_{\text{SFT}}}
\log \pi_{\theta}
\big(
\text{COT}_{C}
\mid C
\big),
\end{equation}
where $\pi_{\theta}$ denotes the LLM policy parameterized by $\theta$.

\subsubsection{Policy Optimization}

After supervised fine-tuning, we further refine the model using GRPO. During this stage, the model takes as input an anchor node and a candidate set of nodes, and generates multiple candidate multi-hop paths. These generated paths are evaluated using a reward function, which guides the model to produce socially plausible and cross-community connections.

\paragraph{Path Length Preference}
To discourage degenerate single-hop predictions, we reward paths whose length approaches the maximum hop limit
\begin{equation}
R_{\text{len}} = \frac{|C|-1}{6}.
\end{equation}

\paragraph{Social Homophily}
To preserve semantic coherence, we encourage nodes along the path to remain close to the anchor user in the profile embedding space
\begin{equation}
R_{\text{homo}}
=
\frac{1}{|C|}
\sum_{i=0}^{|C|-1}
\cos(\mathbf{x}_{v_0}, \mathbf{x}_{v_i}).
\end{equation}

\paragraph{Influence-aware Traversal}
To promote shortcut formation across communities, we reward paths that traverse structurally influential nodes
\begin{equation}
R_{\text{inf}}
=
\frac{1}{|C|-2}
\sum_{i=1}^{|C|-2}
\tilde{d}^{-}(v_i).
\end{equation}

The final reward for GRPO is computed as
\begin{equation}
R_{\text{GSI}}
=
R_{\text{len}} + R_{\text{homo}} + R_{\text{inf}},
\end{equation}
which guides the model to generate socially coherent multi-hop connections during training.

\section{GraphMind-Botnet}
In this section, we introduce {GraphMind-Botnet}, a simulated social network constructed by autonomous GraphMind agents. The resulting network exhibits structural properties that closely resemble those of real-world human social graphs, including community organization and small-world connectivity.

\subsection{Basic Framework}

\paragraph{User Roles.}
GraphMind agents are assigned user roles to simulate realistic online profiles and individual preferences. Usernames and profile descriptions are generated by an LLM and stored in a centralized database. Demographic attributes, including age, gender, education level, and geographic location, are sampled to match the empirical distributions observed in the TwiBot-22 training dataset. This design ensures that the synthetic population reflects realistic user diversity at the attribute level.

\paragraph{MCP-based Interaction Execution.}
All external social actions, such as following, liking, and commenting, are abstracted as callable operations under the Model Context Protocol (MCP)~\cite{2025Model}. By encapsulating environment interactions into standardized MCP interfaces, agents are able to interact with the simulated social platform in a structured and modular manner. This abstraction also facilitates extensibility, allowing additional interaction types or platform rules to be incorporated without modifying the agent logic.

\subsection{Botnet Construction}

GraphMind agents autonomously construct the simulated social network through heuristic reasoning and iterative interaction with the environment. The construction process consists of two tightly coupled components: the generation of follow relationships to establish network topology, and the refinement of interaction behaviors to populate the network with realistic social activity.

\subsubsection{Follow Relationship Generation}

The formation of follow relationships proceeds by first establishing cohesive intra-community structures and subsequently introducing cross-community connections to ensure global connectivity. To this end, bots are partitioned into $N$ communities based on the profile embeddings of LLM-driven agents, such that agents within the same community share similar demographic characteristics and content preferences. Guided by the language-style similarity principle~\cite{2020Language}, bots within each community naturally form dense and cohesive subgraphs, capturing the homophilic tendencies commonly observed in online social networks.

\looseness=-1
Beyond intra-community connectivity, real-world social platforms such as Facebook and X are known to form globally connected graphs that exhibit the six degrees of separation phenomenon~\cite{ugander2011anatomy,myers2014information}. To reproduce this property, GraphMind introduces multi-hop follow relationships between otherwise disconnected agents. Specifically, pairs of nodes without an existing path between them are randomly sampled, and the corresponding bots are prompted to infer plausible multi-hop follow chains based on the current network topology. These inferred paths are then materialized by invoking the network simulation module through MCP, which instantiates the missing follow edges. Repeating this process incrementally yields a structurally complex and well-connected botnet. The overall procedure is briefly summarized in Algorithm~\ref{alg:Multi-hop generation}.

\subsubsection{Interaction Behavior Refinement}

Once the follow network has been established and exhibits small-world characteristics, agents engage in fine-grained social interactions with their one-hop neighbors to populate the network with heterogeneous behaviors. At this stage, each LLM-driven bot analyzes the profile attributes and inferred preferences of its neighbors, estimates the corresponding social tie strength. Then they generate diversified interaction behaviors that are consistent with this relationship. These interactions, including likes, comments, and other observable actions, are executed and recorded through the MCP interface.

Through this two-stage construction process, GraphMind-Botnet evolves into a socially coherent network in which both the structural topology and the interaction patterns closely mirror those of human-operated social platforms.

\begin{table}[t]
    \centering
    \caption{The composition of the our dataset}
    \label{tab:graphmind-dataset}
    \resizebox{\columnwidth}{!}{
    \begin{tabular}{lccccccccc}
    \hline
     \textbf{Human} & \textbf{Bot} & \textbf{Edges} & \textbf{Edge Types} & \textbf{Communities}\\
    \hline
     1,000 & 1000 & 41375 & 5 & 50\\
    \hline
    \end{tabular}
    }
\end{table}

\subsection{GraphMind Dataset}
\paragraph{Composition.} We aggregate the follower network and interaction records generated during simulation and integrate them with human nodes sampled from the TwiBot-20 dataset~\cite{feng2021twibot} to construct the GraphMind Dataset. Table~\ref{tab:graphmind-dataset} summarizes its composition, including the number of human users, bots, edge types, and communities. A visualization of the following network is provided in Appendix. The bot network exhibits both dense intra-community structures and sparse yet meaningful inter-community links. These human-like characteristics enhance the realism of the dataset and increase its potential to confuse detection.

\begin{table*}[!htbp]
    \centering
    \caption{Bot detection performance across datasets (lower values indicate stronger evasion).}
    \label{tab:performance}

    \vspace{-4pt} 

    \tiny 
    \setlength{\tabcolsep}{8pt}
    \renewcommand{\arraystretch}{0.80} 

    \resizebox{\textwidth}{!}{
    \begin{tabular}{lcccccccccccc}
    \toprule
    
    \multirow{2}{*}{Dataset} & \multirow{2}{*}{Metric} & \multicolumn{4}{c}{Metadata-based} & Text-based & \multicolumn{2}{c}{Homo-GNN} & \multicolumn{3}{c}{Heter-GNN} \\
    \cmidrule(lr){3-6} \cmidrule(lr){7-7} \cmidrule(lr){8-9} \cmidrule(lr){10-12}
     & & AB & RF & DT & SVM & Wei \textit{et al.} & GCN & GAT & BotRGCN & RGT & S-HGN \\
    \midrule

    \multirow{2}{*}{Cresci-15} 
     & Acc &  96.9 & 96.0 & 94.7 & 95.4 & 94.78 & 98.2 & 97.7 & 96.3 & 97.2 & 96.9 \\
     & F1 & 95.5 & 95.6 & 94.6 & 95.3 & 84.25 & 96.0 & 97.0 & 96.3 & 96.5 & 95.9 \\
    \midrule
    
    \multirow{2}{*}{Cresci-17} 
     & Acc &  93.2 & 88.1 & 86.2 & 85.1 & 86.30 & / & / & / & / & / \\
     & F1 & 83.4 & 78.9 & 79.4 & 76.8 & 79.40 & / & / & / & / & / \\
    \midrule
    
    \multirow{2}{*}{Twibot-20} 
     & Acc &  85.9 & 85.4 & 81.1 & 85.7 & 78.26 & 76.8 & 83.2 & 86.8 & 87.4 & 85.9 \\
     & F1 & 84.6 & 83.9 & 80.2 & 84.8 & 76.5  & 75.3 & 81.9 & 86.6 & 85.7 & 84.7 \\
    \midrule
    
    \multirow{2}{*}{MGTAB-22} 
     & Acc &  92.5 & 90.1 & 88.1 & 87.7 & / & 85.2 & 87.4 & 89.6 & 92.1 & 90.4 \\
     & F1 & 88.6 & 87.8 & 85.7 & 86.3 & / & 78.8 & 84.3 & 87.2 & 90.4 & 87.7 \\
    \midrule
    
    \multirow{2}{*}{Twibot-22} 
     & Acc &  67.3 & 73.6 & 74.6 & 78.4 & 69.7 & 81.6 & 76.3 & 81.6 & 73.9 & 76.7 \\
     & F1 & 35.8 & 32.4 & 50.6 & 52.6 & 54.6 & 56.8 & 54.6 & 58.4 & 45.1 & 45.7 \\
    \midrule
    
    \multirow{2}{*}{EvoBot} 
     & Acc &  74.4 & 69.3 & 70.1 & 75.7 & 58.8 & 78.1 & 75.7 & 88.5 & 80.3 & 82.6 \\
     & F1 & 70.3 & 65.7 & 66.6 & 70.8 & 54.6 & 68.4 & 61.5 & 83.2 & 77.8 & 80.3 \\
    \midrule
    
    \multicolumn{12}{c}{{ChatGPT-3.5}} \\
    \midrule
    
    \multirow{2}{*}{OASIS} 
     & Acc &  67.9 & 73.4 & 66.8 & 70.4 & 68.3 & 80.5 & 70.7 & 82.7 & 76.3 & 85.6 \\
     & F1 & 51.6 & 44.5 & 50.6 & 61.3 & 59.2 & 79.8 & 62.5 & 81.3 & 75.6 & 80.5 \\
    \midrule
    
    \multirow{2}{*}{BotSim-24} 
     & Acc &  77.5 & 75.7 & 71.4 & 74.4 & 50.8 & 72.7 & 80.3 & 89.9 & 82.3 & 87.7 \\
     & F1 & 74.8 & 72.4 & 68.5 & 69.8 & 50.4 & 50.5 & 73.1 & 86.7 & 76.4 & 83.1 \\

    \midrule
    \multicolumn{12}{c}{{Qwen3 1.7B}} \\
    \midrule
    
    \multirow{2}{*}{GraphMind Dataset}
    & Acc &  74.4 & \textbf{69.3} & 70.1 & 72.7 & 50.8 & {61.9} & {76.8} & {72.4} & 73.5 & {70.7} \\
    & F1 & 70.3 & 65.7 & 66.6 & 70.8 & 44.6 & 46.5 & 67.3 & 67.6 & 65.8 & 60.5 \\
    \midrule
    \multicolumn{12}{c}{{Qwen3 7B}} \\
    \midrule
    \multirow{2}{*}{GraphMind Dataset} 
    & Acc & \textbf{71.3} & {69.6} & \textbf{65.7} & \textbf{68.4} & \textbf{46.7} & \textbf{60.1} & \textbf{75.7} & \textbf{70.5} & \textbf{69.6} & \textbf{69.1} \\
    & F1  & 68.8 & 66.6 & 63.0 & 64.2 & 46.4 & 58.4 & 61.5 & 62.6 & 65.3 & 68.4 \\
    \bottomrule
    \end{tabular}
    }
    \footnotesize
    \textit{Note: Lower values correspond to better evasion performance and reduced detectability.}
\end{table*}

\paragraph{Model and Training.} We adopt Qwen3 1.7B and Qwen3 8B~\cite{yang2025qwen3} as the base LLMs for GraphMind agents. Both models provide a favorable trade-off between model capacity and computational efficiency, enabling scalable simulation of large social networks while retaining sufficient expressive power to capture complex social behaviors. Training and inference are performed on a single NVIDIA RTX 4090 GPU, with each network generation requiring approximately 18 hours. Both modules of the GraphMind framework are trained using datasets of 3k samples. The current GraphMind Dataset is generated based on training and inference using the Qwen-3 1.7B model. Detailed model architectures and training hyperparameters are provided in Appendix~D.

\section{Experiment}
\subsection{Experimental Settings}
\begin{figure*}[!ht]
    \centering
    \begin{subfigure}[t]{0.32\linewidth}
        \centering
        \includegraphics[width=\linewidth]{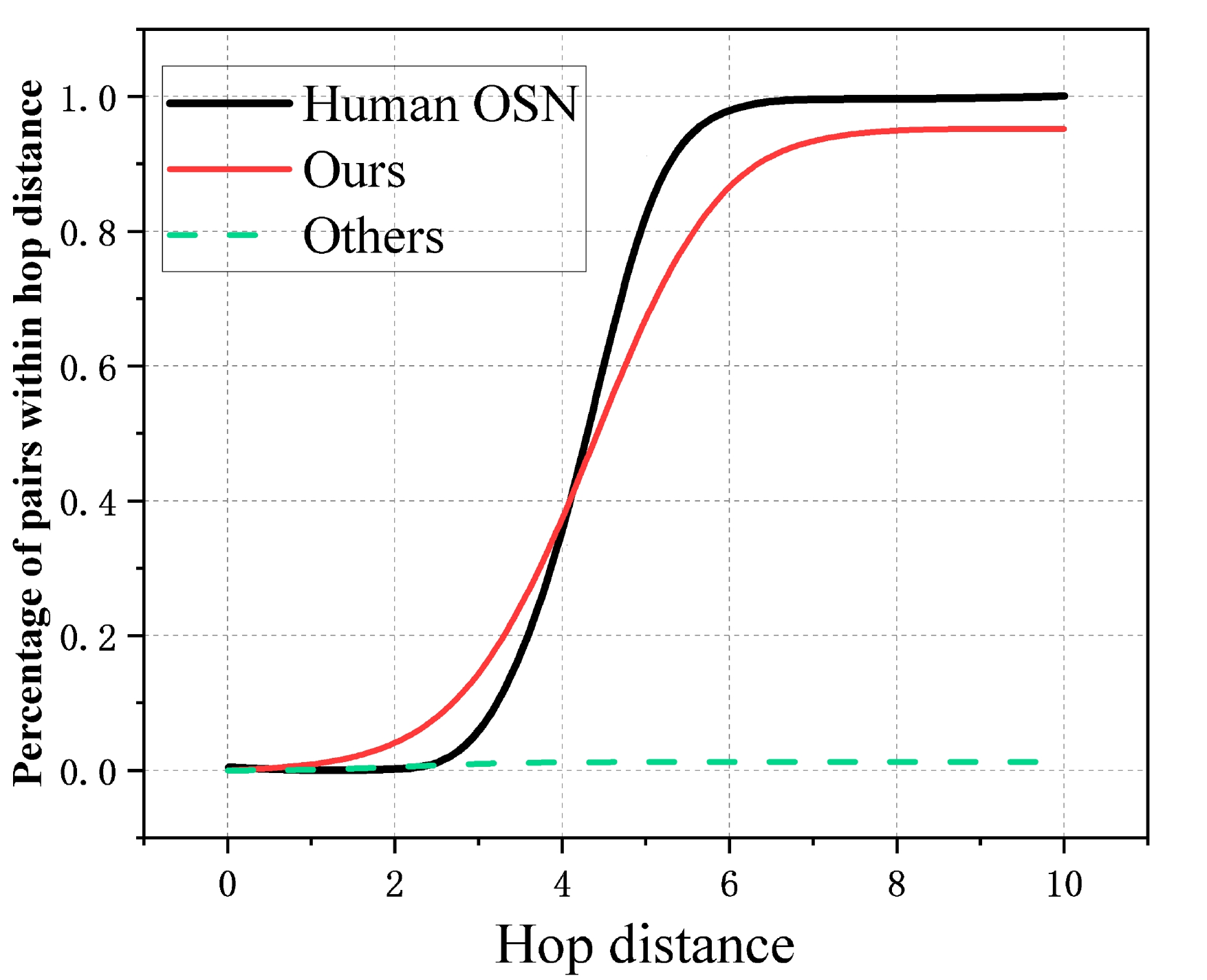}
        \caption{Hop Distances}
        \label{fig:hop-distance}
    \end{subfigure}
    \hspace{-5pt}
    \begin{subfigure}[t]{0.31\linewidth}
        \centering
        \includegraphics[width=\linewidth]{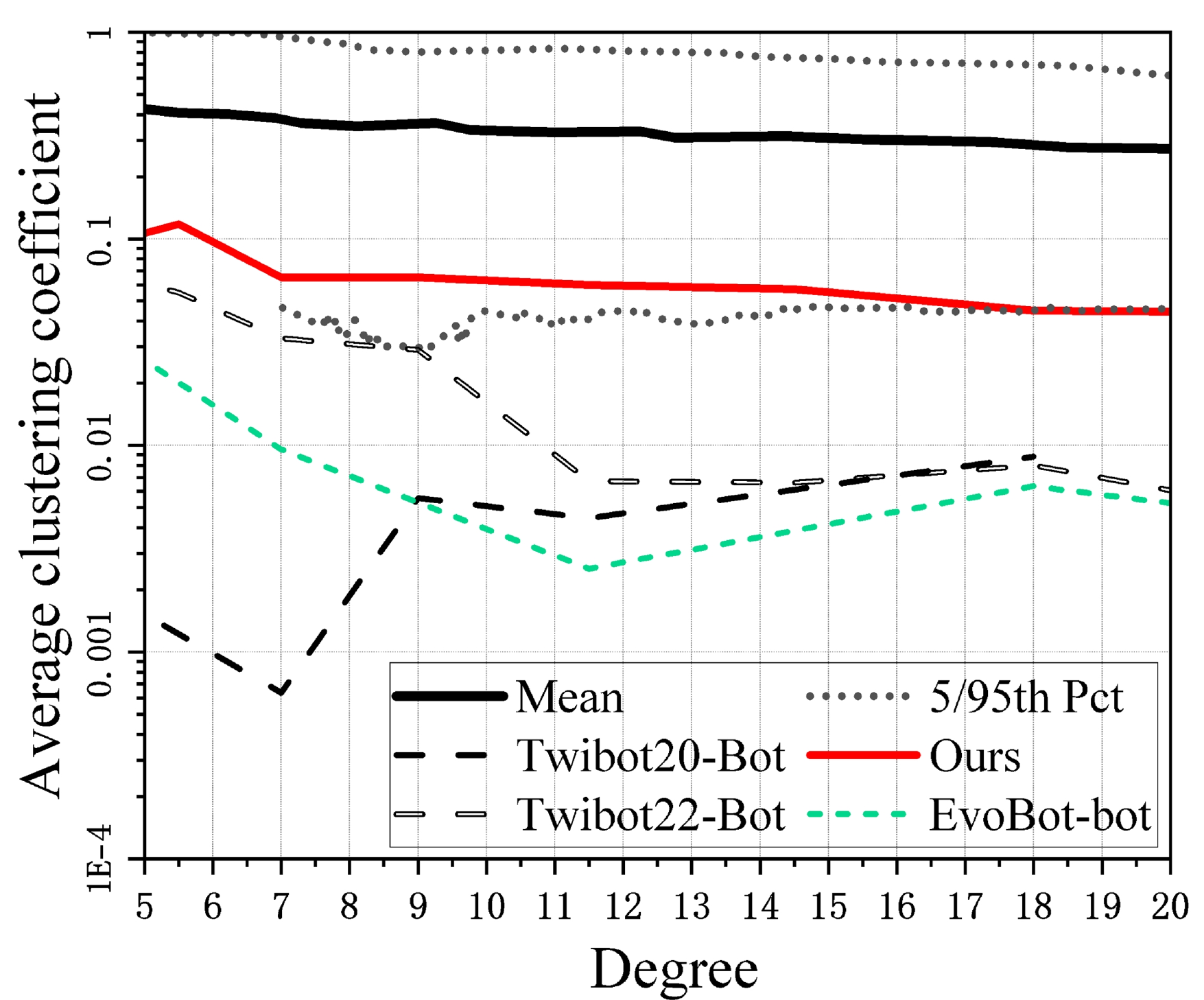}
        \caption{Clustering Coefficient}
        \label{fig:clustering-coef}
    \end{subfigure}
    \begin{subfigure}[t]{0.32\linewidth}
        \centering
        \includegraphics[width=\linewidth]{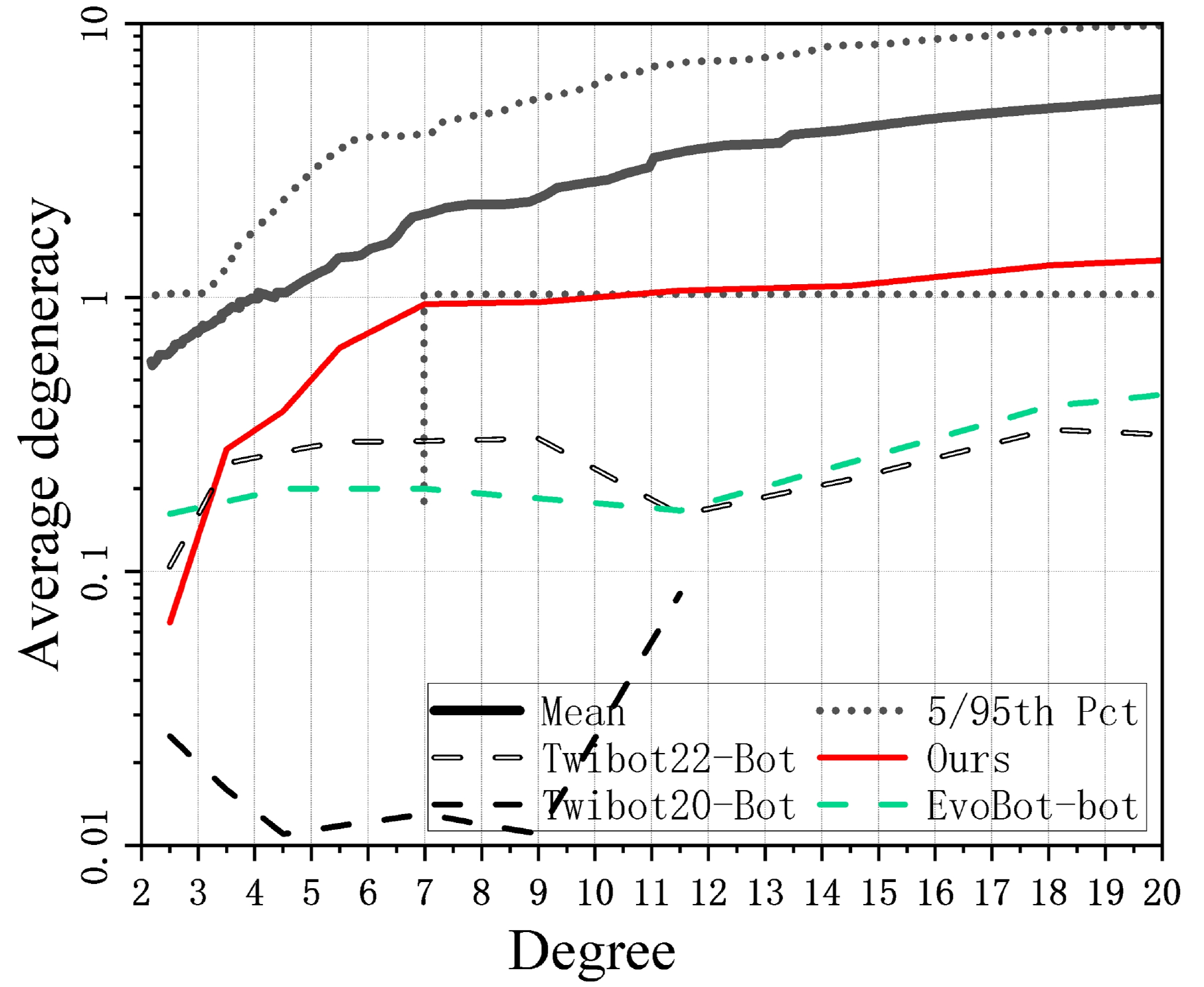}
        \caption{Average Degeneracy}
        \label{fig:degeneracy}
    \end{subfigure}

    \caption{Structural property analysis of different network}
    \label{fig:osn-structure}
\end{figure*}

\begin{figure}[t]
    \centering
    \setlength{\abovecaptionskip}{4pt}
    \includegraphics[width=\columnwidth]{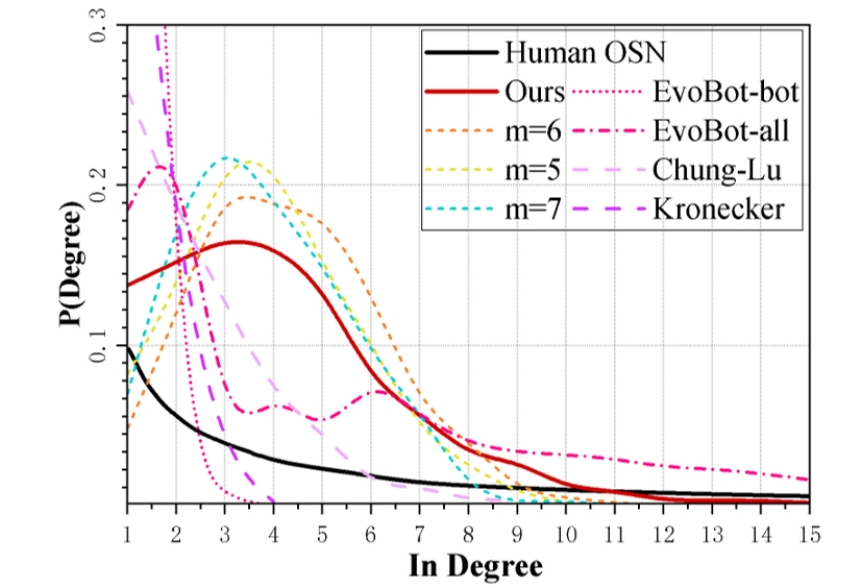}
    \caption{Degree distributions}
    \label{fig:In-Degree}
\end{figure}
\paragraph{Datasets.} 
We evaluate the effectiveness of the GraphMind Dataset by comparing it against existing social bot datasets and simulation frameworks. Traditional benchmarks, including Cresci-15~\cite{cresci2015fame}, Cresci-17~\cite{cresci2017paradigm}, TwiBot-20~\cite{feng2021twibot}, TwiBot-22~\cite{fengtwibot}, and MGTAB-22~\cite{shi2025mgtab}, serve as references for performance comparison.
To fairly reproduce other LLM-driven social bot simulations, we follow the original authors’ open-source implementations: OASIS~\cite{yang2024oasis} is simulated with 1,000 agents over 50 interaction rounds, whereas BotSim~\cite{qiao2025botsim} and EvoBot~\cite{kong2025enhancing} directly use the datasets released by the original works. Since OASIS only contains bot samples, we supplement it with 1,000 human nodes to ensure a comparable human–bot mixture with GraphMind Dataset.

\paragraph{Detectors.} 
We consider four categories of detection algorithms: 
(1) Meta-based detectors, including AdaBoost~\cite{zhu2009multi}, Random Forest~\cite{yang2020scalable}, Decision Tree (DT)~\cite{lepping2018wiley}, and SVM~\cite{boser1992training}; 
(2) Text-based detectors~\cite{wei2019twitter}; 
(3) Homogeneous GNN detectors, such as GCN~\cite{kipf2016semi} and GAT~\cite{velivckovic2017graph}; and 
(4) Heterogeneous GNN detectors, including S-HGN~\cite{lv2021we}, BotRGCN~\cite{feng2021botrgcn}, and RGT~\cite{feng2022twibot}. This setup allows a thorough evaluation of both traditional and LLM-driven botnets under diverse detection settings.
Detection models are trained using an 80/20 split on the respective training set. Detectors are trained on TwiBot-22 and evaluated on each LLM-driven social bot simulation (OASIS, BotSim, EvoBot, and GraphMind).
This cross-dataset setting is deliberately adopted to quantify the detectability gap between simulated bot networks and real human networks.

\subsection{Experiment Results}
\subsubsection{Bot Detection Evasion Performance} 
Table~\ref{tab:performance} reports the evasion performance of GraphMind social bot against a wide range of bot detection methods. 
Since lower detection accuracy or F1 score corresponds to stronger evasion, the results show that GraphMind dataset consistently degrades the effectiveness of existing detectors. Notably, both homogeneous and heterogeneous GNN-based models suffer substantial performance drops (F1 scores of 60-70\% compared to 85-95\% on traditional datasets), indicating that GraphMind social bots present significant challenges for structure-aware detectors. While detection remains feasible, the explicit modeling of network topology substantially increases the difficulty of bot identification.
This observation suggests that explicitly enhancing network-level realism in LLM-driven social simulations significantly increases the challenge of bot detection.
Notably, even with a lightweight base model (Qwen3-1.7B), GraphMind dataset reduces GNN detection accuracy to 61.9\% (F1: 46.5\%), representing a substantial improvement in evasion capability compared to prior LLM-based botnet simulations (OASIS: 80.5\% acc, BotSim: 72.7\% acc).

\subsubsection{Network Structure Analysis} 
To further analyze the differences between our botnet, human networks, and prior methods, we qualitatively evaluate their structural commonalities using standard network analysis metrics.

\paragraph{Path Length.}
The distribution of shortest-path distances between node pairs is a fundamental macroscopic property of social networks.
Following prior work on OSNs~\cite{ugander2011anatomy}, we analyze follow-path lengths and compare our method with existing social bot simulation approaches.
We characterize node-to-node proximity using the neighborhood function $P(h)$, defined as the fraction of node pairs $(u, v)$ such that $u$ is reachable from $v$ by a path of length at most $h$. 
As shown in Figure~\ref{fig:hop-distance}, human social networks exhibit an average pairwise distance of 4.7, with approximately 92\% of user pairs connected within five hops~\cite{ugander2011anatomy}.
Our generated botnet closely matches this empirical pattern, achieving an average hop distance of 4.77, with 90\% of node pairs connected within six hops.

In contrast, existing social bot simulation methods, including OASIS~\cite{yang2024oasis}, EvoBot~\cite{kong2025enhancing}, and BotSim~\cite{qiao2025botsim}, produce extremely sparse graphs with poor global connectivity.
Most nodes in these baselines remain isolated, resulting in overall reachability below 2\% and preventing meaningful path-length statistics.

\paragraph{Clustering Coefficient and Degeneracy.} 
We evaluate the structural realism of generated social graphs using the local clustering coefficient and graph degeneracy, which capture local cohesion and densely connected cores. Following prior work~\cite{leskovec2008statistical}, directed follow edges are treated as undirected via a Facebook-style neighbor projection. Figure~\ref{fig:clustering-coef} plots the average clustering coefficient as a function of node degree, with the empirical mean and 5th/95th percentiles of Facebook networks as reference. For degrees below 20, human social networks exhibit clustering coefficients above 30\%, and only GraphMind falls within this empirical range, while other botnets deviate substantially.

We further analyze graph degeneracy, defined as the maximum non-empty $k$-core. As shown in Figure~\ref{fig:degeneracy}, GraphMind consistently lies within the percentile range. 
Botnets from TwiBot20/22 and recent LLM-driven simulations such as OASIS~\cite{yang2024oasis} and BotSim~\cite{qiao2025botsim} yield near-zero values on both metrics, indicating fragmented graphs with limited core structure. All metrics are computed primarily for nodes with degree $<20$, which is appropriate for small synthetic graphs and enables reliable characterization of local structural patterns.

\begin{figure*}[!ht]
    \centering
    \begin{subfigure}[t]{0.32\linewidth}
        \centering
        \setlength{\fboxrule}{0.1pt}
        \setlength{\fboxsep}{0pt}
        \fcolorbox{gray!60}{white}{%
            \includegraphics[width=\linewidth]{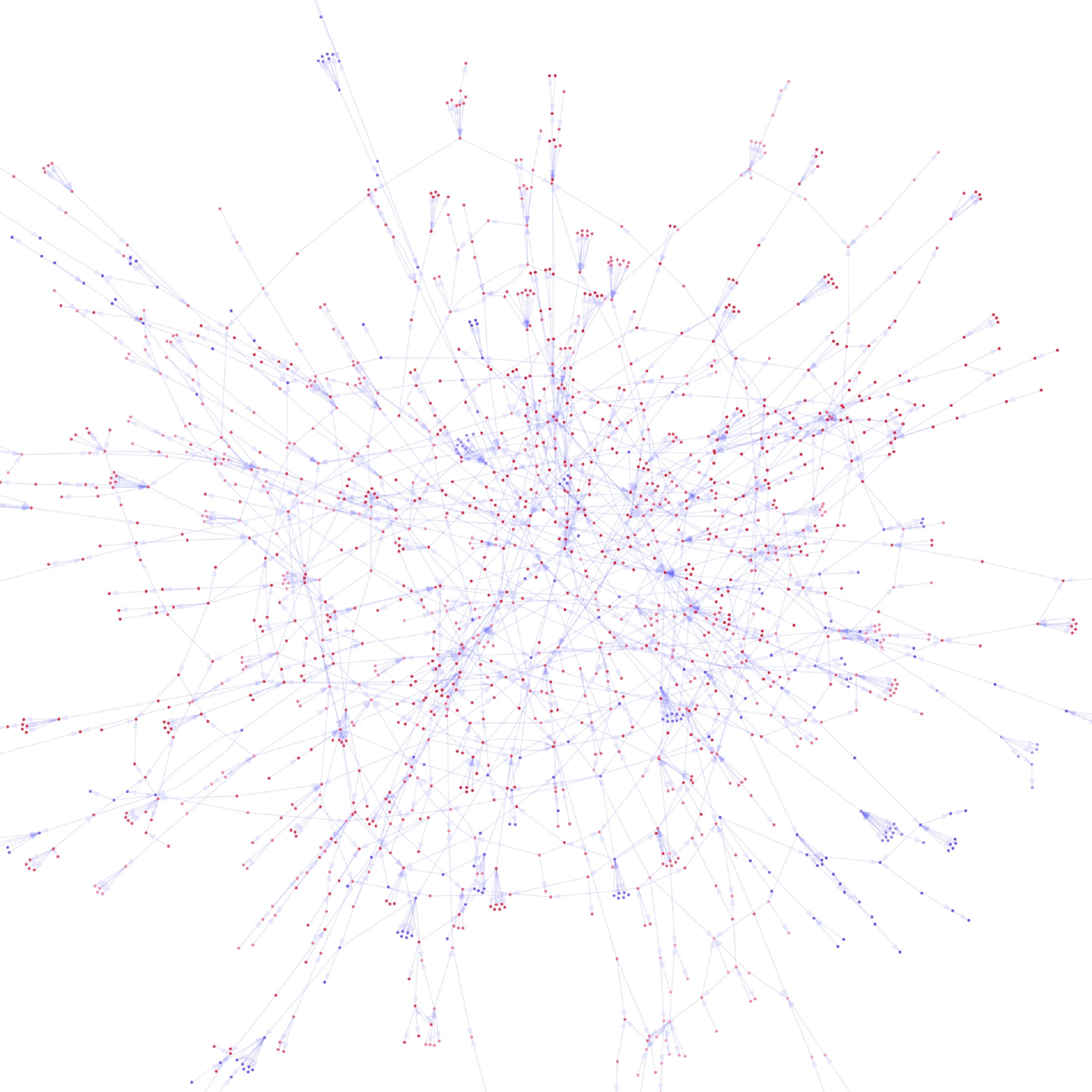}
        }
        \caption{Visualization of human follow network in TwiBot20}
        \label{fig:twibot20-human(part)}
    \end{subfigure}
    \hfill
    \begin{subfigure}[t]{0.32\linewidth}
        \centering
        \setlength{\fboxrule}{0.1pt}
        \setlength{\fboxsep}{0pt}
        \fcolorbox{gray!60}{white}{%
            \includegraphics[width=\linewidth]{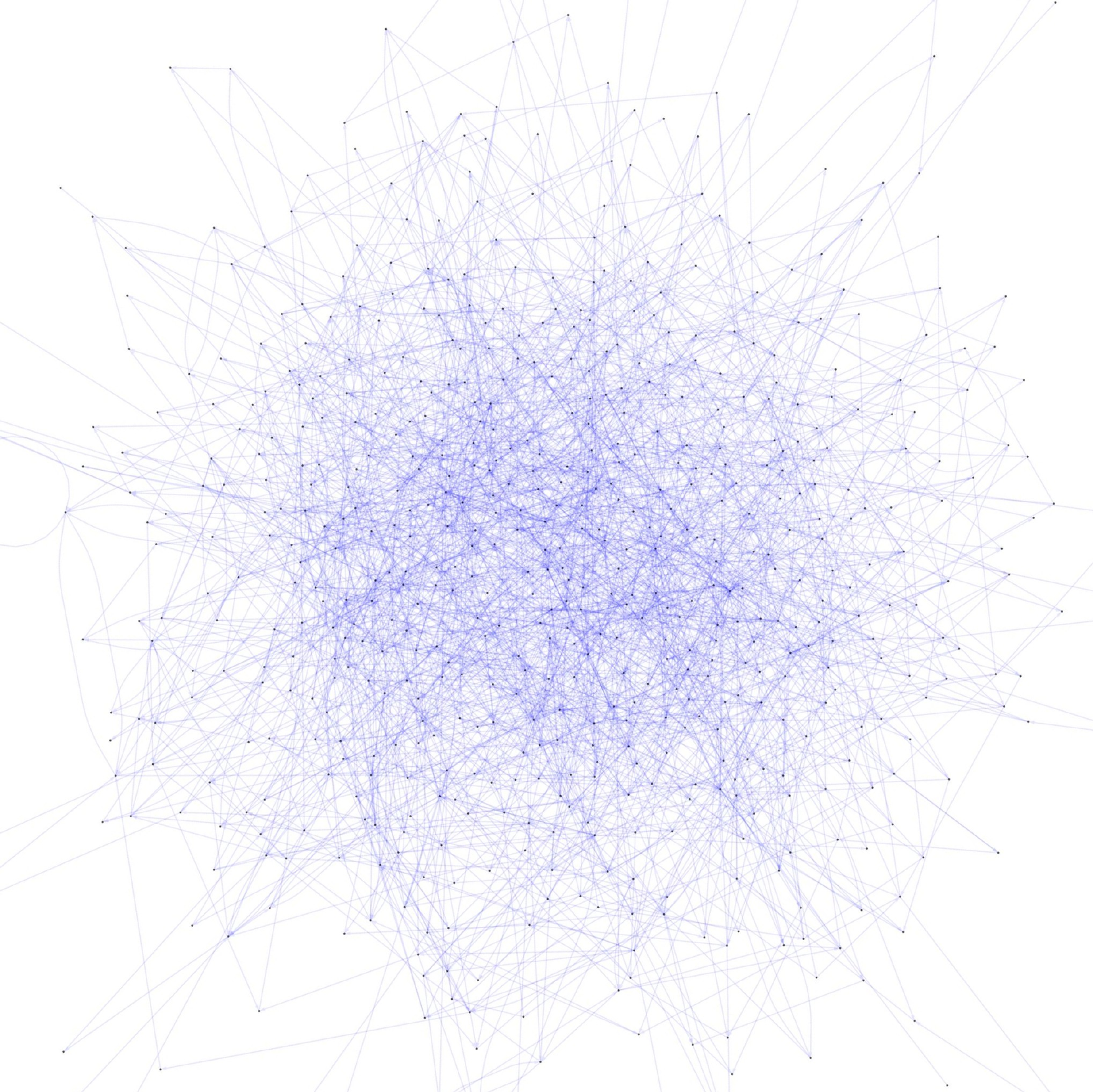}
        }
        \caption{Visualization of bot follow network in GraphMind (ours)}
        \label{fig:graphmind}
    \end{subfigure}
    \hfill
    \begin{subfigure}[t]{0.32\linewidth}
        \centering
        \setlength{\fboxrule}{0.1pt}
        \setlength{\fboxsep}{0pt}
        \fcolorbox{gray!60}{white}{%
            \includegraphics[width=\linewidth]{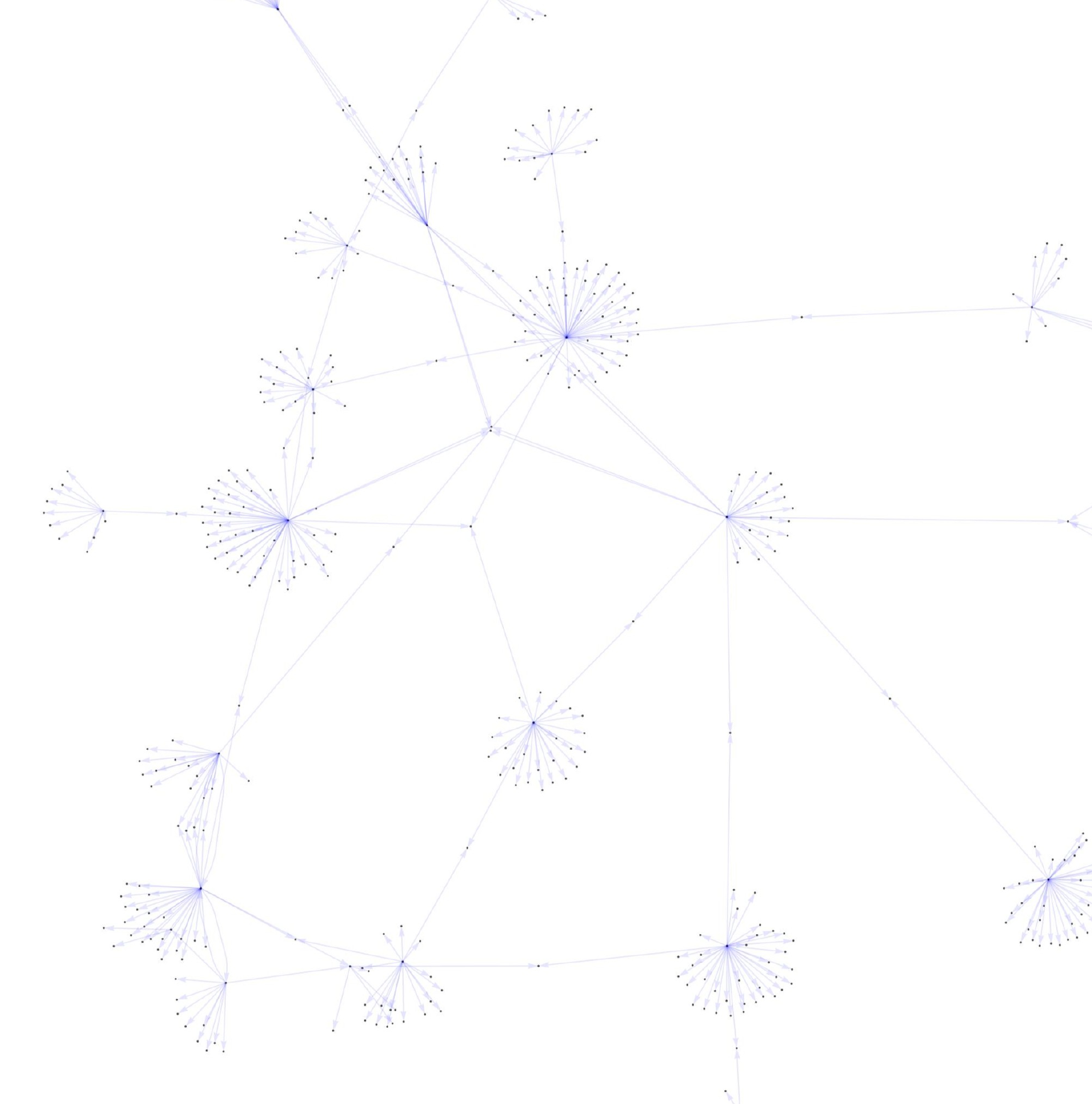}
        }
        \caption{Visualization of bot follow network in Evobot (part)}
        \label{fig:Evobotnet}
    \end{subfigure}

    \caption{Network visualizations comparing human and bot follow networks across different datasets.}
    \label{fig:visualizations}
\end{figure*}

\paragraph{Network visualizations.}
To further validate the accuracy of our data analysis, we visualize several representative botnet and human follow networks. Figures~\ref{fig:twibot20-human(part)} present the human follow network in TwiBot20, which exhibits a mesh-like structure characterized by a densely connected core and progressively sparser peripheral regions. In contrast, existing social bot simulations predominantly form a star-like topology, with a single dominant core and a large number of isolated nodes, as shown in Figure~\ref{fig:Evobotnet}. By comparison, the network constructed by our method more closely resembles the mesh-like topology observed in human networks (Figure~\ref{fig:graphmind}), introducing a substantial number of plausible follow edges and resulting in a large, well-connected component rather than fragmented network structures.

\begin{table}[t]
    \centering
    \caption{Ablation Study of GraphMind Modules on Detection Evasion (\%).}
    \label{tab:ablation_study}
    \resizebox{\columnwidth}{!}{
    \begin{tabular}{lcccccc}
        \toprule
        \multirow{2}{*}{\textbf{Methods}} 
        & \multicolumn{2}{c}{\textbf{S-HGN}}
        & \multicolumn{2}{c}{\textbf{BotRGCN}}
        & \multicolumn{2}{c}{\textbf{RGT}}\\
        \cmidrule(lr){2-3} \cmidrule(lr){4-5} \cmidrule(lr){6-7} 
        & Acc & F1 & Acc & F1 & Acc & F1 \\
        \midrule
        \rowcolor[gray]{0.9} {All} 
        & & & & & & \\
        \quad Qwen3-1.7B & 70.7 & 60.5 & 72.4 & 67.6 & 73.5 & 65.8 \\
        \quad Qwen3-8B   & 69.1 & 68.4 & 70.5 & 62.6 & 69.6 & 65.3 \\
        \rowcolor[gray]{0.9} \textbf{w/o FIM} 
        & & & & & & \\
        \quad Qwen3-1.7B & 78.9 & 69.5 & 79.8 & 74.2 & 80.5 & 73.1 \\
        \quad Qwen3-8B   & 77.2 & 75.1 & 78.4 & 70.5 & 77.9 & 72.8 \\
        \rowcolor[gray]{0.9} \textbf{w/o FIM+GSI} 
        & & & & & & \\
        \quad Qwen3-1.7B & 86.4 & 80.2 & 87.5 & 83.4 & 88.1 & 81.0 \\
        \quad Qwen3-8B   & 85.8 & 84.5 & 86.2 & 81.1 & 85.5 & 82.3 \\
        \bottomrule
    \end{tabular}
    }
    \footnotesize
    \textit{Note: \textbf{FIM} denotes Fine-Grained Interaction Modeling, \textbf{GSI} refers to Graph-Augmented Social Inference. Lower Accuracy and F1 indicate stronger bot evasion.}
\end{table}


        



\subsection{Experimental Analysis}
\subsubsection{Ablation Experiments}
The fine-grained interaction modeling module enables a richer action repertoire in bot–bot interactions.
As shown by the ablation study in Table~\ref{tab:ablation_study}, removing this component leads to a substantial increase in detection accuracy for heterogeneous GNN-based methods. This is because heterogeneous GNNs are designed to capture diversity in node-level operations, and the absence of fine-grained interaction modeling causes the botnet to degenerate structurally. Specifically, agents exhibit action homogeneity, repeatedly executing a limited set of operations (e.g., commenting), which makes the network easier to detect.
In contrast, agents trained with this module dynamically select context-appropriate actions conditioned on tweet intent, resulting in behavior sequences that exhibit stronger contextual coherence and emotional consistency.

\subsubsection{Comparison against simple graph generators}
To compare multi-hop connections generated by LLM-driven agents with randomly constructed ones, we conduct a controlled experiment in which communities are connected either by agent inference or by random $m$-hop chains, where $m$ denotes the path length. 
In the random setting, disconnected node pairs are repeatedly sampled and linked via $m-2$ randomly selected intermediate nodes until 97\% of node pairs become reachable.
We evaluate the resulting graphs using the in-degree distribution (Figure~\ref{fig:In-Degree}). 
Compared to random strategies, LLMs tend to prioritize and identify influential core nodes within communities, characterized by high in-degree centrality.
As a result, the GraphMind-Botnet exhibits a heavy-tailed, power-law-like in-degree distribution with several high-impact nodes~\cite{myers2014information}.
Figure~\ref{fig:In-Degree} also includes comparisons with classical graph generators, including the Chung–Lu model~\cite{fasino2021generating} and the Kronecker graph model~\cite{leskovec2010kronecker}. GraphMind simultaneously preserves realistic long-tail behavior and uniformity consistent with real social networks. Notably, although the Evobo dataset appears long-tailed at the aggregate level, isolating bot nodes reveals that most bots have in-degree one and none exceed degree four, indicating a superficial rather than structural long-tail effect.
Additional visualizations and statistics are provided in Appendix~E.


\section{Conclusion}
GraphMind pioneers seamless integration of social graphs with LLM-powered bots via dual-stage training, enabling autonomous human-like topology construction. The GNN-Augmented Social Inference module incorporates small-world effects and homophily principles for multi-hop connection discovery, while the Fine-Grained Interaction Modeling module decouples behavioral operations from semantic generation to implement affectively coherent atomic actions via MCP protocol. The resultant botnet exhibits topological properties that closely resemble those of human networks, underscoring the need for resilient countermeasures against evolving LLM social bots.


\bibliography{custom}

@article{shi2025mgtab,
  title={Mgtab: A multi-relational graph-based twitter account detection benchmark},
  author={Shi, Shuhao and Qiao, Kai and Liu, Zihao and Yang, Jie and Chen, Chen and Chen, Jian and Yan, Bin},
  journal={Neurocomputing},
  pages={130490},
  year={2025},
  publisher={Elsevier}
}

@article{leskovec2010kronecker,
  title={Kronecker graphs: an approach to modeling networks.},
  author={Leskovec, Jure and Chakrabarti, Deepayan and Kleinberg, Jon and Faloutsos, Christos and Ghahramani, Zoubin},
  journal={Journal of Machine Learning Research},
  volume={11},
  number={2},
  year={2010}
}

@article{fasino2021generating,
  title={Generating large scale-free networks with the Chung--Lu random graph model},
  author={Fasino, Dario and Tonetto, Arianna and Tudisco, Francesco},
  journal={Networks},
  volume={78},
  number={2},
  pages={174--187},
  year={2021},
  publisher={Wiley Online Library}
}

@article{hu2022lora,
  title={Lora: Low-rank adaptation of large language models.},
  author={Shen, Yelong and Wallis, Phillip and Allen-Zhu, Zeyuan and Li, Yuanzhi and Wang, Shean and others}
}

@inproceedings{myers2014information,
  title={Information network or social network? The structure of the Twitter follow graph},
  author={Myers, Seth A and Sharma, Aneesh and Gupta, Pankaj and Lin, Jimmy},
  booktitle={Proceedings of the 23rd international conference on world wide web},
  pages={493--498},
  year={2014}
}

@inproceedings{park2023generative,
  title={Generative agents: Interactive simulacra of human behavior},
  author={Park, Joon Sung and O'Brien, Joseph and Cai, Carrie Jun and Morris, Meredith Ringel and Liang, Percy and Bernstein, Michael S},
  booktitle={Proceedings of the 36th annual acm symposium on user interface software and technology},
  pages={1--22},
  year={2023}
}

@article{arnaboldi2016ego,
  title={Ego network structure in online social networks and its impact on information diffusion},
  author={Arnaboldi, Valerio and Conti, Marco and La Gala, Massimiliano and Passarella, Andrea and Pezzoni, Fabio},
  journal={Computer Communications},
  volume={76},
  pages={26--41},
  year={2016},
  publisher={Elsevier}
}

@article{ugander2011anatomy,
  title={The anatomy of the facebook social graph},
  author={Ugander, Johan and Karrer, Brian and Backstrom, Lars and Marlow, Cameron},
  journal={arXiv preprint arXiv:1111.4503},
  year={2011}
}

@article{yang2024oasis,
  title={Oasis: Open agent social interaction simulations with one million agents},
  author={Yang, Ziyi and Zhang, Zaibin and Zheng, Zirui and Jiang, Yuxian and Gan, Ziyue and Wang, Zhiyu and Ling, Zijian and Chen, Jinsong and Ma, Martz and Dong, Bowen and others},
  journal={arXiv preprint arXiv:2411.11581},
  year={2024}
}

@article{tabassum2018social,
  title={Social network analysis: An overview},
  author={Tabassum, Shazia and Pereira, Fabiola SF and Fernandes, Sofia and Gama, Jo{\~a}o},
  journal={Wiley Interdisciplinary Reviews: Data Mining and Knowledge Discovery},
  volume={8},
  number={5},
  pages={e1256},
  year={2018},
  publisher={Wiley Online Library}
}

@article{bei2025graphs,
  title={Graphs Meet AI Agents: Taxonomy, Progress, and Future Opportunities},
  author={Bei, Yuanchen and Zhang, Weizhi and Wang, Siwen and Chen, Weizhi and Zhou, Sheng and Chen, Hao and Li, Yong and Bu, Jiajun and Pan, Shirui and Yu, Yizhou and others},
  journal={arXiv preprint arXiv:2506.18019},
  year={2025}
}

@article{wu2025llm,
  title={Llm fine-tuning: Concepts, opportunities, and challenges},
  author={Wu, Xiao-Kun and Chen, Min and Li, Wanyi and Wang, Rui and Lu, Limeng and Liu, Jia and Hwang, Kai and Hao, Yixue and Pan, Yanru and Meng, Qingguo and others},
  journal={Big Data and Cognitive Computing},
  volume={9},
  number={4},
  pages={87},
  year={2025},
  publisher={MDPI}
}

@inproceedings{leskovec2008statistical,
  title={Statistical properties of community structure in large social and information networks},
  author={Leskovec, Jure and Lang, Kevin J and Dasgupta, Anirban and Mahoney, Michael W},
  booktitle={Proceedings of the 17th international conference on World Wide Web},
  pages={695--704},
  year={2008}
}

@inproceedings{devalal2018lora,
  title={LoRa technology-an overview},
  author={Devalal, Shilpa and Karthikeyan, A},
  booktitle={2018 second international conference on electronics, communication and aerospace technology (ICECA)},
  pages={284--290},
  year={2018},
  organization={IEEE}
}

@article{ekin2023prompt,
  title={Prompt engineering for ChatGPT: a quick guide to techniques, tips, and best practices},
  author={Ekin, Sabit},
  journal={Authorea Preprints},
  year={2023},
  publisher={Authorea}
}

@article{kong2025enhancing,
  title={Enhancing llm-based social bot via an adversarial learning framework},
  author={Kong, Fanqi and Zhang, Xiaoyuan and Chen, Xinyu and Yang, Yaodong and Zhu, Song-Chun and Feng, Xue},
  journal={arXiv preprint arXiv:2508.17711},
  year={2025}
}

@inproceedings{feng2021twibot,
  title={Twibot-20: A comprehensive twitter bot detection benchmark},
  author={Feng, Shangbin and Wan, Herun and Wang, Ningnan and Li, Jundong and Luo, Minnan},
  booktitle={Proceedings of the 30th ACM international conference on information \& knowledge management},
  pages={4485--4494},
  year={2021}
}

@article{cresci2015fame,
  title={Fame for sale: Efficient detection of fake Twitter followers},
  author={Cresci, Stefano and Di Pietro, Roberto and Petrocchi, Marinella and Spognardi, Angelo and Tesconi, Maurizio},
  journal={Decision Support Systems},
  volume={80},
  pages={56--71},
  year={2015},
  publisher={Elsevier}
}

@inproceedings{cresci2017paradigm,
  title={The paradigm-shift of social spambots: Evidence, theories, and tools for the arms race},
  author={Cresci, Stefano and Di Pietro, Roberto and Petrocchi, Marinella and Spognardi, Angelo and Tesconi, Maurizio},
  booktitle={Proceedings of the 26th international conference on world wide web companion},
  pages={963--972},
  year={2017}
}

@misc{lepping2018wiley,
  title={Wiley Interdisciplinary Reviews: Data Mining and Knowledge Discovery},
  author={Lepping, Joachim},
  year={2018}
}

@inproceedings{feng2021satar,
  title={Satar: A self-supervised approach to twitter account representation learning and its application in bot detection},
  author={Feng, Shangbin and Wan, Herun and Wang, Ningnan and Li, Jundong and Luo, Minnan},
  booktitle={Proceedings of the 30th ACM international conference on information \& knowledge management},
  pages={3808--3817},
  year={2021}
}

@inproceedings{ali2019detect,
  title={Detect me if you can: Spam bot detection using inductive representation learning},
  author={Ali Alhosseini, Seyed and Bin Tareaf, Raad and Najafi, Pejman and Meinel, Christoph},
  booktitle={Companion proceedings of the 2019 world wide web conference},
  pages={148--153},
  year={2019}
}

@article{zhou2023sotopia,
  title={Sotopia: Interactive evaluation for social intelligence in language agents},
  author={Zhou, Xuhui and Zhu, Hao and Mathur, Leena and Zhang, Ruohong and Yu, Haofei and Qi, Zhengyang and Morency, Louis-Philippe and Bisk, Yonatan and Fried, Daniel and Neubig, Graham and others},
  journal={arXiv preprint arXiv:2310.11667},
  year={2023}
}

@article{gao2023s3,
  title={S3: Social-network simulation system with large language model-empowered agents},
  author={Gao, Chen and Lan, Xiaochong and Lu, Zhihong and Mao, Jinzhu and Piao, Jinghua and Wang, Huandong and Jin, Depeng and Li, Yong},
  journal={arXiv preprint arXiv:2307.14984},
  year={2023}
}

@inproceedings{park2022social,
  title={Social simulacra: Creating populated prototypes for social computing systems},
  author={Park, Joon Sung and Popowski, Lindsay and Cai, Carrie and Morris, Meredith Ringel and Liang, Percy and Bernstein, Michael S},
  booktitle={Proceedings of the 35th Annual ACM Symposium on User Interface Software and Technology},
  pages={1--18},
  year={2022}
}

@article{zhu2009multi,
  title={Multi-class adaboost},
  author={Zhu, Ji and Zou, Hui and Rosset, Saharon and Hastie, Trevor and others},
  journal={Statistics and its Interface},
  volume={2},
  number={3},
  pages={349--360},
  year={2009}
}

@inproceedings{boser1992training,
  title={A training algorithm for optimal margin classifiers},
  author={Boser, Bernhard E and Guyon, Isabelle M and Vapnik, Vladimir N},
  booktitle={Proceedings of the fifth annual workshop on Computational learning theory},
  pages={144--152},
  year={1992}
}

@article{kipf2016semi,
  title={Semi-Supervised Classification with Graph Convolutional Networks},
  author={Kipf, TN},
  journal={arXiv preprint arXiv:1609.02907},
  year={2016}
}

@article{velivckovic2017graph,
  title={Graph attention networks},
  author={Veli{\v{c}}kovi{\'c}, Petar and Cucurull, Guillem and Casanova, Arantxa and Romero, Adriana and Lio, Pietro and Bengio, Yoshua},
  journal={arXiv preprint arXiv:1710.10903},
  year={2017}
}

@inproceedings{lv2021we,
  title={Are we really making much progress? revisiting, benchmarking and refining heterogeneous graph neural networks},
  author={Lv, Qingsong and Ding, Ming and Liu, Qiang and Chen, Yuxiang and Feng, Wenzheng and He, Siming and Zhou, Chang and Jiang, Jianguo and Dong, Yuxiao and Tang, Jie},
  booktitle={Proceedings of the 27th ACM SIGKDD conference on knowledge discovery \& data mining},
  pages={1150--1160},
  year={2021}
}

@inproceedings{feng2021botrgcn,
  title={BotRGCN: Twitter bot detection with relational graph convolutional networks},
  author={Feng, Shangbin and Wan, Herun and Wang, Ningnan and Luo, Minnan},
  booktitle={Proceedings of the 2021 IEEE/ACM international conference on advances in social networks analysis and mining},
  pages={236--239},
  year={2021}
}

@article{2020Language,
  title={Language-Style Similarity and Social Networks:},
  author={ Kovacs, Balazs  and  Kleinbaum, Adam M },
  journal={Psychol Sci},
  number={2},
  year={2020},
}

@article{2025Model,
  title={Model Context Protocol (MCP): Landscape, Security Threats, and Future Research Directions},
  author={ Hou, Xinyi  and  Zhao, Yanjie  and  Wang, Shenao  and  Wang, Haoyu },
  year={2025},
  journal={ArXiv}
}

@inproceedings{fengtwibot,
  year={2022},
  title={TwiBot-22: Towards Graph-Based Twitter Bot Detection},
  author={Feng, Shangbin and Tan, Zhaoxuan and Wan, Herun and Wang, Ningnan and Chen, Zilong and Zhang, Binchi and Zheng, Qinghua and Zhang, Wenqian and Lei, Zhenyu and Yang, Shujie and others},
  booktitle={Thirty-sixth Conference on Neural Information Processing Systems Datasets and Benchmarks Track}
}

@article{he2024botdgt,
  title={Botdgt: Dynamicity-aware social bot detection with dynamic graph transformers},
  author={He, Buyun and Yang, Yingguang and Wu, Qi and Liu, Hao and Yang, Renyu and Peng, Hao and Wang, Xiang and Liao, Yong and Zhou, Pengyuan},
  journal={arXiv preprint arXiv:2404.15070},
  year={2024}
}

@inproceedings{wu2024botscl,
  title={Botscl: Heterophily-aware social bot detection with supervised contrastive learning},
  author={Wu, Qi and Yang, Yingguang and He, Buyun and Liu, Hao and Yang, Renyu and Liao, Yong},
  booktitle={International Conference on Pattern Recognition},
  pages={53--68},
  year={2024},
  organization={Springer}
}

@inproceedings{yang2020scalable,
  title={Scalable and generalizable social bot detection through data selection},
  author={Yang, Kai-Cheng and Varol, Onur and Hui, Pik-Mai and Menczer, Filippo},
  booktitle={Proceedings of the AAAI conference on artificial intelligence},
  volume={34},
  number={01},
  pages={1096--1103},
  year={2020}
}

@inproceedings{wei2019twitter,
  title={Twitter bot detection using bidirectional long short-term memory neural networks and word embeddings},
  author={Wei, Feng and Nguyen, Uyen Trang},
  booktitle={2019 First IEEE International conference on trust, privacy and security in intelligent systems and applications (TPS-ISA)},
  pages={101--109},
  year={2019},
  organization={IEEE}
}

@article{feng2022twibot,
  title={Twibot-22: Towards graph-based twitter bot detection},
  author={Feng, Shangbin and Tan, Zhaoxuan and Wan, Herun and Wang, Ningnan and Chen, Zilong and Zhang, Binchi and Zheng, Qinghua and Zhang, Wenqian and Lei, Zhenyu and Yang, Shujie and others},
  journal={Advances in Neural Information Processing Systems},
  volume={35},
  pages={35254--35269},
  year={2022}
}

@inproceedings{qiao2025botsim,
  title={BotSim: LLM-Powered Malicious Social Botnet Simulation},
  author={Qiao, Boyu and Li, Kun and Zhou, Wei and Li, Shilong and Lu, Qianqian and Hu, Songlin},
  booktitle={Proceedings of the AAAI Conference on Artificial Intelligence},
  volume={39},
  number={13},
  pages={14377--14385},
  year={2025}
}

@article{tan2023walklm,
  title={Walklm: A uniform language model fine-tuning framework for attributed graph embedding},
  author={Tan, Yanchao and Zhou, Zihao and Lv, Hang and Liu, Weiming and Yang, Carl},
  journal={Advances in neural information processing systems},
  volume={36},
  pages={13308--13325},
  year={2023}
}

@article{milgram1967small,
  title={The small world problem},
  author={Milgram, Stanley and others},
  journal={Psychology today},
  volume={2},
  number={1},
  pages={60--67},
  year={1967},
  publisher={New York}
}

@article{yang2025qwen3,
  title={Qwen3 technical report},
  author={Yang, An and Li, Anfeng and Yang, Baosong and Zhang, Beichen and Hui, Binyuan and Zheng, Bo and Yu, Bowen and Gao, Chang and Huang, Chengen and Lv, Chenxu and others},
  journal={arXiv preprint arXiv:2505.09388},
  year={2025}
}

@article{wei2022chain,
  title={Chain-of-thought prompting elicits reasoning in large language models},
  author={Wei, Jason and Wang, Xuezhi and Schuurmans, Dale and Bosma, Maarten and Xia, Fei and Chi, Ed and Le, Quoc V and Zhou, Denny and others},
  journal={Advances in neural information processing systems},
  volume={35},
  pages={24824--24837},
  year={2022}
}

@article{shao2024deepseekmath,
  title={Deepseekmath: Pushing the limits of mathematical reasoning in open language models},
  author={Shao, Zhihong and Wang, Peiyi and Zhu, Qihao and Xu, Runxin and Song, Junxiao and Bi, Xiao and Zhang, Haowei and Zhang, Mingchuan and Li, YK and Wu, Yang and others},
  journal={arXiv preprint arXiv:2402.03300},
  year={2024}
}

\newpage
\appendix
\section{Related work}
\subsection{Social Simulation Based on LLM.}
Simulation of human behavior based on large language models (LLMs) has recently emerged as a prominent research frontier. 
The Smallville sandbox world enables natural language interaction with a town composed of twenty-five agents, demonstrating both believable individual behaviors and emergent social dynamics \cite{park2023generative}. 
SimReddit introduces an LLM-driven simulation platform to study intentional social interactions \cite{park2022social}. 
S³ \cite{gao2023s3} combines Markov chains with LLMs to model opinion dynamics, while SOTOPIA proposes a framework for evaluating social intelligence \cite{zhou2023sotopia}.
Moving toward online social network (OSN) simulation, 
OASIS implements a general and scalable social media simulator to investigate large-scale collective phenomena and behaviors~\cite{yang2024oasis}.
BotSim designs an LLM-driven malicious social bot network simulator~\cite{qiao2025botsim}, 
and EvoBot bypasses co-evolving detection systems through human-like expression strategies~\cite{kong2025enhancing}.

Despite these advances, existing frameworks primarily focus on refining LLMs’ ability to imitate individual-level actions, while largely overlooking the explicit modeling of*social links between agents, such as follow or friendship relations. Under large-scale social simulation settings, such relational structures are typically assumed or manually specified, which is neither realistic nor scalable.

\subsection{GNN-based Bot Detection}
Graph-based detection methods have demonstrated strong effectiveness against bot evasion, particularly on benchmark datasets that incorporate relational graph information.
Early work introduced Graph Convolutional Networks (GCNs) for bot detection by jointly modeling node attributes and network topology \cite{ali2019detect}. Subsequent studies proposed self-supervised GCN-based frameworks to enhance representation learning \cite{feng2021satar}, as well as relational GCN architectures that support multi-relational aggregation in heterogeneous social graphs \cite{feng2021botrgcn}.

However, these methods become increasingly inadequate as LLM-based social simulations evolve sophisticated evasion capabilities through authentic behavioral emulation and network construction, necessitating novel detection algorithms for advanced LLM-driven tactics.

\section{Limitations}
Current approaches exhibit several limitations. 
First, existing network generation methods struggle to scale to large-scale social networks. Due to the inherent context length constraints of large language models, when the number of agents exceeds approximately 2k, LLM-based social bots can no longer access complete global network information, leading to distorted or inconsistent link generation.

Second, current methods primarily operate at a macroscopic level by training LLMs to heuristically complete missing social links, rather than maintaining social relationships through fine-grained, agent-level interactions. In future work, we plan to enable LLM-driven social bots to actively establish and maintain accurate social connections at the micro level through continuous local interactions.

Finally, the evaluation of how closely agent-generated networks resemble real human social networks remains largely qualitative. More rigorous quantitative analyses are required to assess structural fidelity, which is crucial for studying information diffusion and large-scale collective behavior in simulated social systems.

\section{Ethics Statement}
Following the original data usage terms, we collect and process data from the publicly available TwiBot-22 and TwiBot-20 datasets. To mitigate privacy risks, we employ prompt-based large language models to remove personally identifiable information from the original text, such as phone numbers and email addresses.

Similar to other LLM -based social simulation frameworks, GraphMind agents and the resulting GraphMind-Botnet can facilitate research on information diffusion and collective behavior by enabling the construction of more realistic social networks. However, the high structural fidelity of such agent-generated networks to real human social networks also introduces potential misuse risks, as they could be exploited by malicious actors to construct deceptive social botnets. To address this concern, we implement a strict application and review process for code and model parameter access, ensuring that all usage is limited to legitimate research purposes.

While GraphMind demonstrates the potential to evade existing social bot detection methods, particularly graph neural network (GNN)-based approaches, we emphasize that our work focuses on foundational modeling rather than deployment. The ethical implications of such capabilities must be carefully considered. As part of future work, we plan to investigate next-generation social bot detection mechanisms that are better suited to identifying LLM-based social bots and coordinated botnets. One promising direction is to leverage large language models to analyze and decompose suspicious network structures: although a botnet may internally control its relational topology, it often remains structurally isolated from the broader social graph. Properties such as the construction mechanisms and volume of follow edges may provide useful signals for botnet-level detection.

These efforts are essential for establishing ethical guidelines, safeguards, and regulatory frameworks that mitigate potential risks. Future research should prioritize the development of transparency and protection mechanisms to ensure responsible use while supporting the formulation of robust governance standards.

\section{Method Details}
GraphMind has the potential to advance human–AI interaction by enabling more realistic social network simulation, with downstream benefits for applications such as collective behavior modeling and social bot detection. 
In addition, the development of detection algorithms capable of accurately identifying LLM-driven social bots will play a critical role in distinguishing human-generated from machine-generated content, thereby supporting the responsible deployment of such technologies. 
Below, we provide additional methodological details that are not explicitly described in the main body of the paper.

\subsection{Fine-Grained Interaction Modeling}
\begin{table}[ht]
    \centering
    \caption{Relationship Circle Definitions}
    \label{tab:rel_circles}
    \begin{tabular}{>{\centering\arraybackslash}p{0.25\linewidth}>{\centering\arraybackslash}p{0.55\linewidth}}
        \toprule
        \textbf{Level} & \textbf{Definition} \\
        \midrule
        \multirow{2}{*}{\makecell{Level 1 \\ (Support clique)}} & Strongest ties; \\
        & frequent contact \\
        \midrule
        \multirow{2}{*}{\makecell{Level 2 \\ (Sympathy group)}} & Close ties; frequent but \\
        & not weekly contact \\
        \midrule
        \multirow{2}{*}{\makecell{Level 3 \\ (Affinity group)}} & Casual ties; \\
        & occasional contact \\
        \midrule
        \multirow{2}{*}{\makecell{Level 4 \\ (Active network)}} & All active ties; \\
        & at least yearly contact \\
        \bottomrule
    \end{tabular}
\end{table}
\subsubsection{Social strength}
In our framework, relationship strength does not directly prescribe specific interaction types, but instead constrains the frequency distribution of different interaction behaviors. This design is inspired by ego network theory, which characterizes strong and weak ties in terms of interaction intensity, frequency, and social cost, rather than specific action semantics. Accordingly, high-strength relationships are associated with more frequent high-engagement actions (e.g., reposting or replying), whereas lower-strength or asymmetric relationships tend to favor low-cost interactions such as liking. We view this mapping as an operationalization of established ego network principles, translating abstract tie strength into observable interaction patterns on online social platforms.

Table~\ref{tab:rel_circles} summarizes our definition of relationship strength. Specifically, we first annotate samples in the dataset by jointly considering interaction frequency between node pairs and profile similarity. During simulation, LLM-driven social bots assign target nodes to discrete relationship strength levels based on this definition, and subsequently generate interaction samples that follow the corresponding behavior frequency distributions.

\subsubsection{Dataset and Prompt}
We sample 3,000 node pairs with existing follow relationships from the TwiBot-22 dataset. Based on their historical interaction frequencies, we annotate the social relationship strength of one node with respect to the other. We then employ DeepSeek to complete the corresponding reasoning process for each annotated pair. Representative examples are shown in Table~\ref{tab:fim-template}.

\subsubsection{Training setting}
\begin{figure}[t]
    \centering
    \includegraphics[width=0.8\columnwidth]{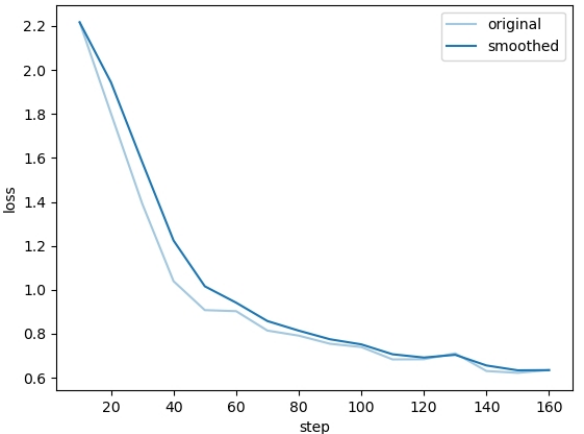}
    \caption{Loss function of FIM}
    \label{fig:loss_fim}
\end{figure}
Since large language models are not pretrained with explicit supervision for this task, we follow a widely adopted industry paradigm and apply supervised fine-tuning (SFT) to initialize the model. This cold-start stage provides stable behavior grounding before subsequent optimization. The training hyperparameters are reported in Table~\ref{tab:sft-fim}. The loss function are shown in Figure~\ref{fig:loss_fim}.
\begin{table}[ht]
    \centering
    \begin{tabular}{l c}
        \hline
        \textbf{Hyperparameter} & \textbf{Value} \\
        \hline
        Optimizer & AdamW \\
        Learning rate & $5 \times 10^{-5}$ \\
        Training epochs & 5 \\
        LR scheduler & Cosine \\
        Warmup steps & 100 \\
        Random seed & 42 \\
        \hline
    \end{tabular}
    \caption{Key hyperparameters used for FIM.}
    \label{tab:sft-fim}
\end{table}

\subsection{Graph-Augmented Social Inference}

\subsubsection{Dataset and Prompt}
We sample 3,000 seed nodes from the TwiBot-22 dataset and collect multi-hop social chains via random walks. For each pair of adjacent nodes, we leverage DeepSeek to infer the underlying rationale for the existence of follow relationships based on node attributes and structural features such as in-degree. The resulting step-by-step reasoning traces are then concatenated into long chains and used for training and optimization, with representative examples shown in Table~\ref{tab:graph-augmented-prompt}.

\subsubsection{Training setting}
\begin{figure}[t]
    \centering
    \includegraphics[width=0.8\columnwidth]{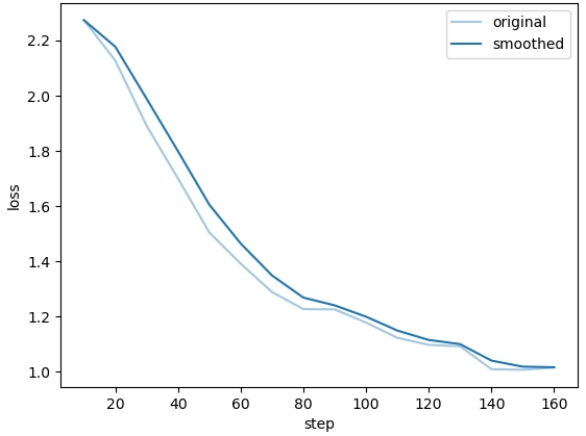}
    \caption{Loss function of GSI}
    \label{fig:loss_gmi}
\end{figure}

The training hyperparameters are reported in Table~\ref{tab:sft-gmi}. The loss function are shown in Figure~\ref{fig:loss_gmi}.

\begin{table}[ht]
    \centering
    \begin{tabular}{l c}
        \hline
        \textbf{Hyperparameter} & \textbf{Value} \\
        \hline
        Optimizer & AdamW \\
        Learning rate & $5 \times 10^{-5}$ \\
        Training epochs & 5 \\
        LR scheduler & Cosine \\
        Warmup steps & 100 \\
        Random seed & 42 \\
        \hline
    \end{tabular}
    \caption{Key hyperparameters used for GSI.}
    \label{tab:sft-gmi}
\end{table}

\begin{table*}[t]
    \centering
    \small
    \begin{tabular}{p{3cm} p{10.5cm}}
    \toprule
    \textbf{Field} & \textbf{Content} \\
    \midrule

    Instruction &
    You are an expert in social media behavior analysis.
    Your task is to infer the social relationship level between two users from an ego-network perspective,
    and then generate a list of interaction actions that follow realistic behavioral patterns.
    Follow the steps below: \newline
    Step 1: Infer the most plausible social relationship level within the ego network. \newline
    Step 2: Based on the inferred relationship strength, determine the expected interaction intensity and cost. \\

    \midrule

    Input &
    \texttt{<user\_profile>}: \{PROFILE\} \newline
    \texttt{<tweets>}: \{TWEETS\} \\

    \midrule

    Think &
    The target user belongs to my ego network with relationship level \{RELATIONSHIP\_LEVEL\}.
    Based on this relationship strength and the observed profile and interaction context,
    I infer appropriate interaction intensity and engagement patterns. \\

    \midrule

    Output &
    A list of interaction actions in the following structured format: \newline
    \texttt{<action>} \newline
    \quad \texttt{<type> like / retweet / comment / follow </type>} \newline
    \quad \texttt{<tweet\_id> \{TWEET\_ID\} </tweet\_id>} \\

    \bottomrule
    \end{tabular}
    \caption{Prompt template used for FIM.}
    \label{tab:fim-template}
\end{table*}

\begin{table*}[t]
    \centering
    \small
    \begin{tabular}{p{3cm} p{10.5cm}}
    \toprule
    \textbf{Field} & \textbf{Content} \\
    \midrule

    Instruction &
    You are an expert in social media behavior analysis. \newline
    Given a source node and a target node, generate a multi-hop following path based on the input social network structure.
    First, select a set of potential intermediate nodes $\{n\}$ from the given network, and then predict and generate potential social links according to their profile information. \newline
    Follow the steps below: \newline
    Step 1: Select candidate intermediate nodes from the input graph based on profile similarity to the source node. \newline
    Step 2: Sequentially infer plausible following relationships between adjacent nodes to form a multi-hop path. \newline
    \newline
    Requirements: \newline
    1. The path must contain $\geq 3$ hops (Source $\rightarrow$ Mediator$_1$ $\rightarrow$ Mediator$_2$ $\rightarrow$ Target). \newline
    2. All mediator nodes must be selected from the given network information. \newline
    3. Path inference must be based on feature similarities (bio, age, education, location, etc.). \newline
    4. The connection rationale must be analyzed for each hop. \\

    \midrule

    Input &
    \texttt{<graph>} \newline
    \quad \texttt{<nodes> \{users\} </nodes>} \newline
    \quad \texttt{<edges> \{follow\_edges\} </edges>} \newline
    \texttt{</graph>} \newline
    \texttt{<source\_node> \{user\_A\} </source\_node>} \newline
    \texttt{<target\_node> \{user\_B\} </target\_node>} \\

    \midrule

    Think &
    \texttt{<think>} \newline
    \texttt{[Link 1]} \newline
    \texttt{<user>1</user> follow <user>2</user>} \newline
    \texttt{<reason>...</reason>} \newline
    \texttt{[Link ..]} \newline
    \texttt{</think>} \\

    \midrule

    Output &
    Based on the analysis of user profiles and social relationships, the most plausible multi-hop following path is: \newline
    \texttt{user\_A $\rightarrow$ mediator\_1 $\rightarrow$ mediator\_2 $\rightarrow$ user\_B}. \\

    \bottomrule
    \end{tabular}
    \caption{Prompt template used for GSI.}
    \label{tab:graph-augmented-prompt}
\end{table*}

\subsection{GraphMind Dataset}
For this dataset, we generate a botnet based on GraphMind agent, following the algorithm described below.
\begin{algorithm}[ht]
    \caption{Multi-hop Follow Completion}
    \label{alg:Multi-hop generation}
    \begin{algorithmic}[1]
    \REQUIRE $G=\{V,E,X,Y\}$; connectivity threshold $\tau$
    \ENSURE $E$
    
    \WHILE{NetworkConnectivity$(G)<\tau$}
        \STATE Sample $(u,v)$ with no path between $u$ and $v$ in $G$
        \STATE $C_u \leftarrow \text{Bot}(G,\;V_{\text{source}}=u,\;V_{\text{target}}=v)$
        \STATE $C_v \leftarrow \text{Bot}(G,\;V_{\text{source}}=v,\;V_{\text{target}}=u)$
        \FOR{each $C\in\{C_u,C_v\}$ with $C=[v_{h_0},\dots,v_{h_k}]$}
            \FOR{$i=0$ \TO $k-1$}
                \STATE $e\leftarrow(v_{h_i}\rightarrow v_{h_{i+1}})$
                \IF{$e\notin E$} \STATE $E\leftarrow E\cup\{e\}$ 
                \ENDIF
            \ENDFOR
        \ENDFOR
    \ENDWHILE
    
    \RETURN $E$
    \end{algorithmic}

\end{algorithm}

\clearpage

\clearpage


\begin{figure*}[t]
    \centering
    \includegraphics[width=0.95\textwidth,height=0.82\textheight,keepaspectratio]{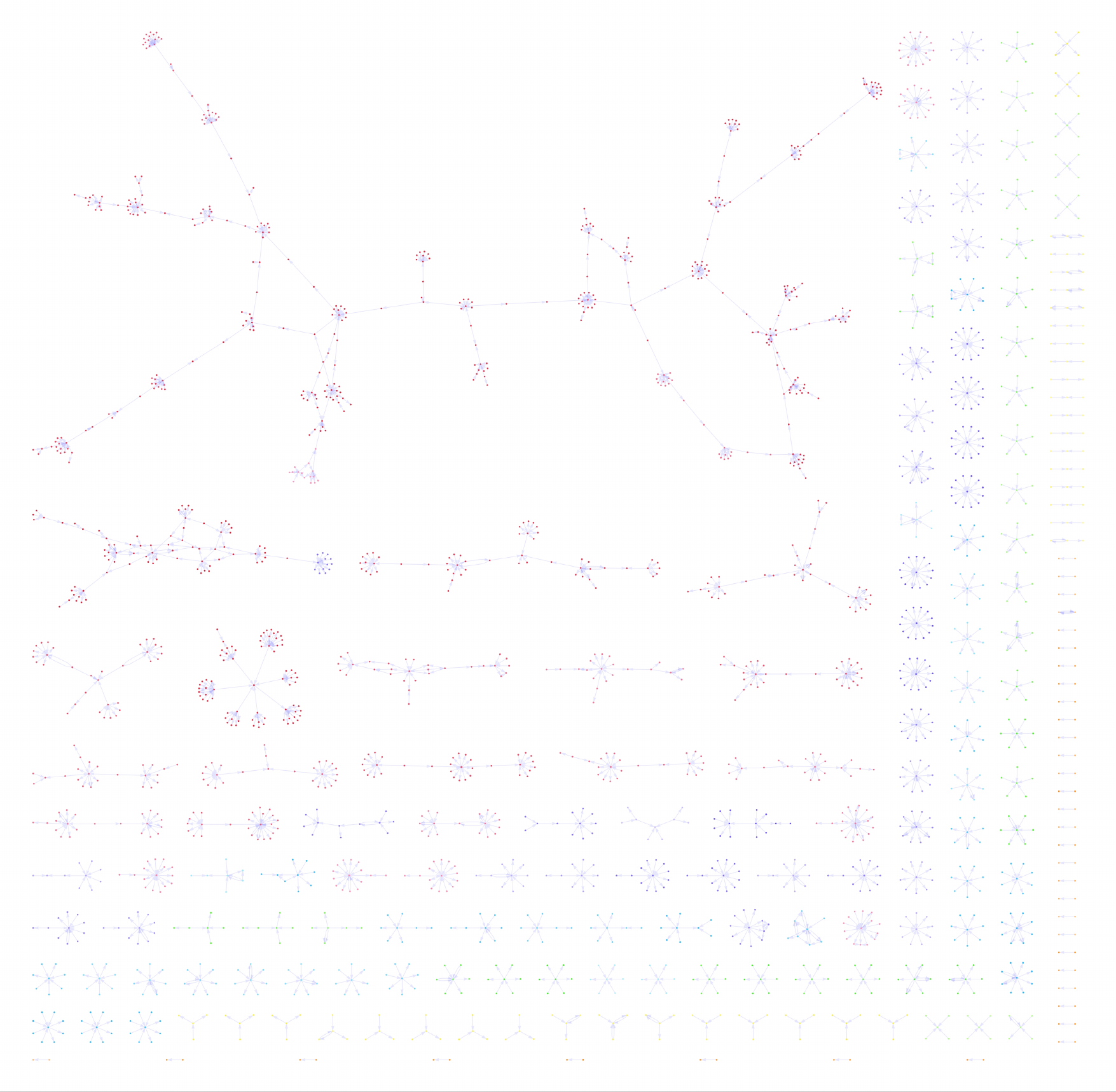}
    \caption{Botnet visualization of Twibot-20.}
    \label{fig:twibot20-bot}
\end{figure*}

\clearpage 

\begin{figure*}[t]
    \centering
    \includegraphics[width=0.95\textwidth,keepaspectratio]{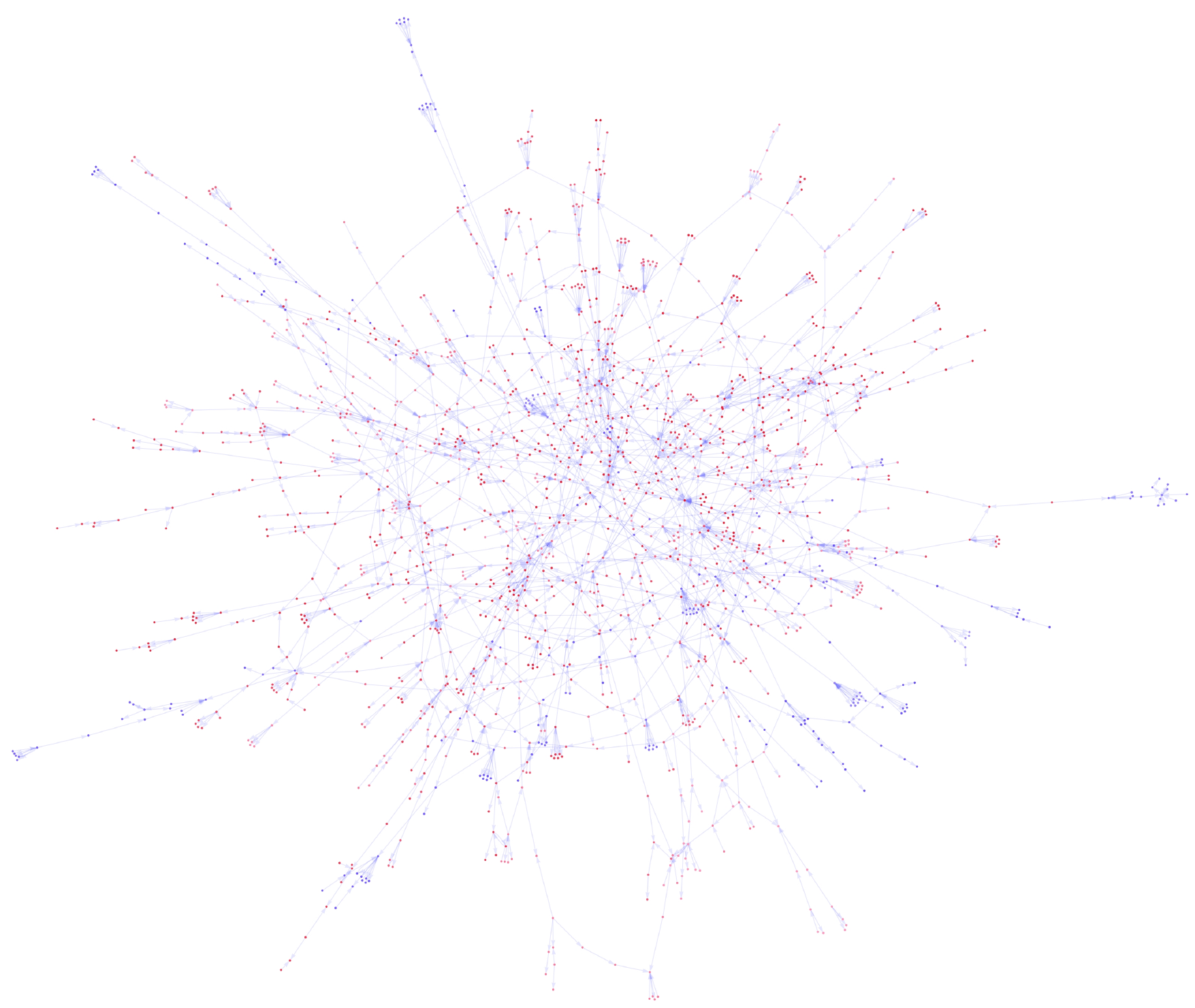}
    \caption{Human network visualization of Twibot-20.}
    \label{fig:twibot20-human}
\end{figure*}

\begin{figure*}[t]
    \centering
    \includegraphics[width=0.95\textwidth,keepaspectratio]{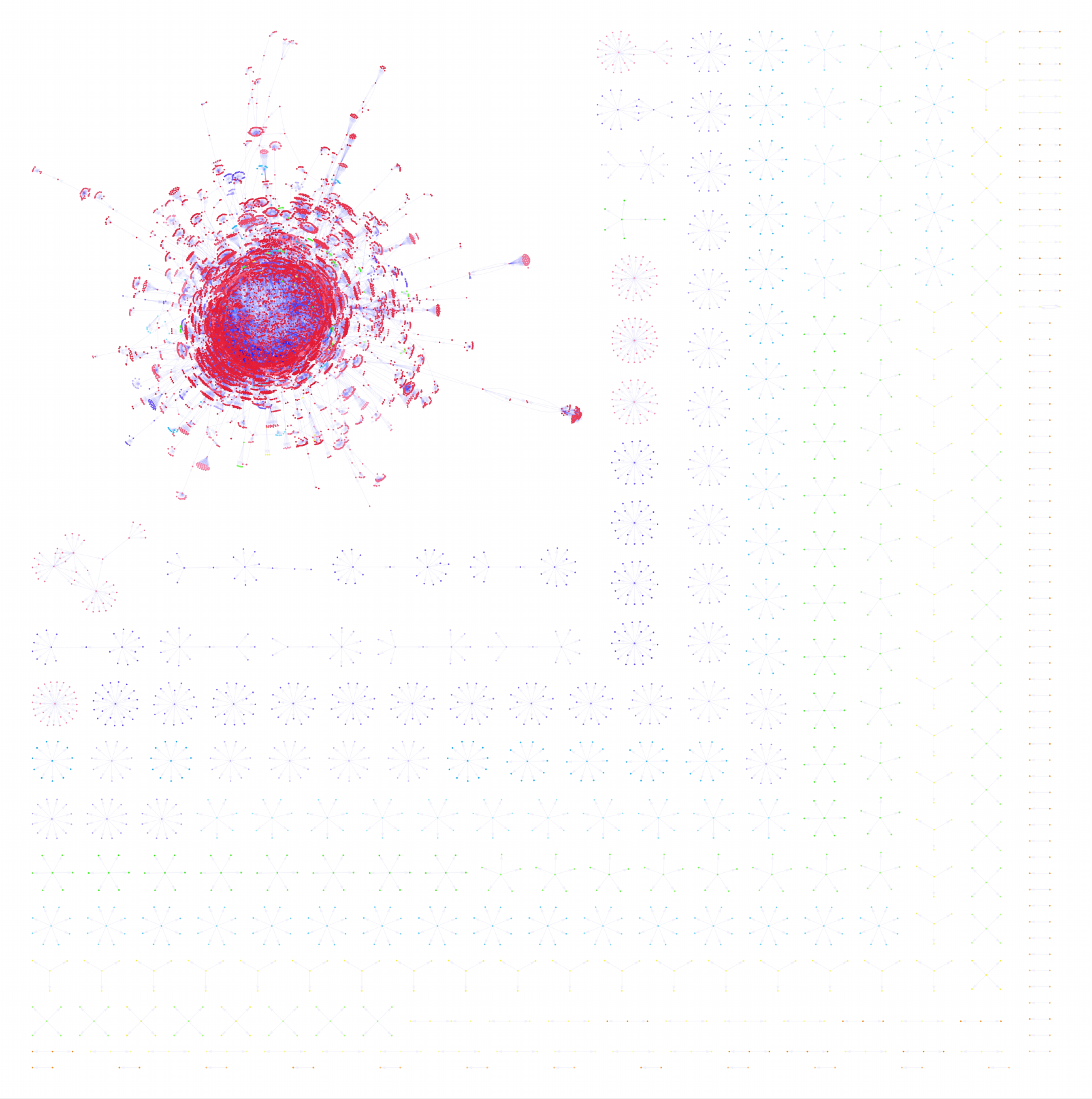}
    \caption{Botnet visualization of Twibot-22.}
    \label{fig:twibot22-botnet}
\end{figure*}

\begin{figure*}[t]
    \centering
    \includegraphics[width=0.95\textwidth,keepaspectratio]{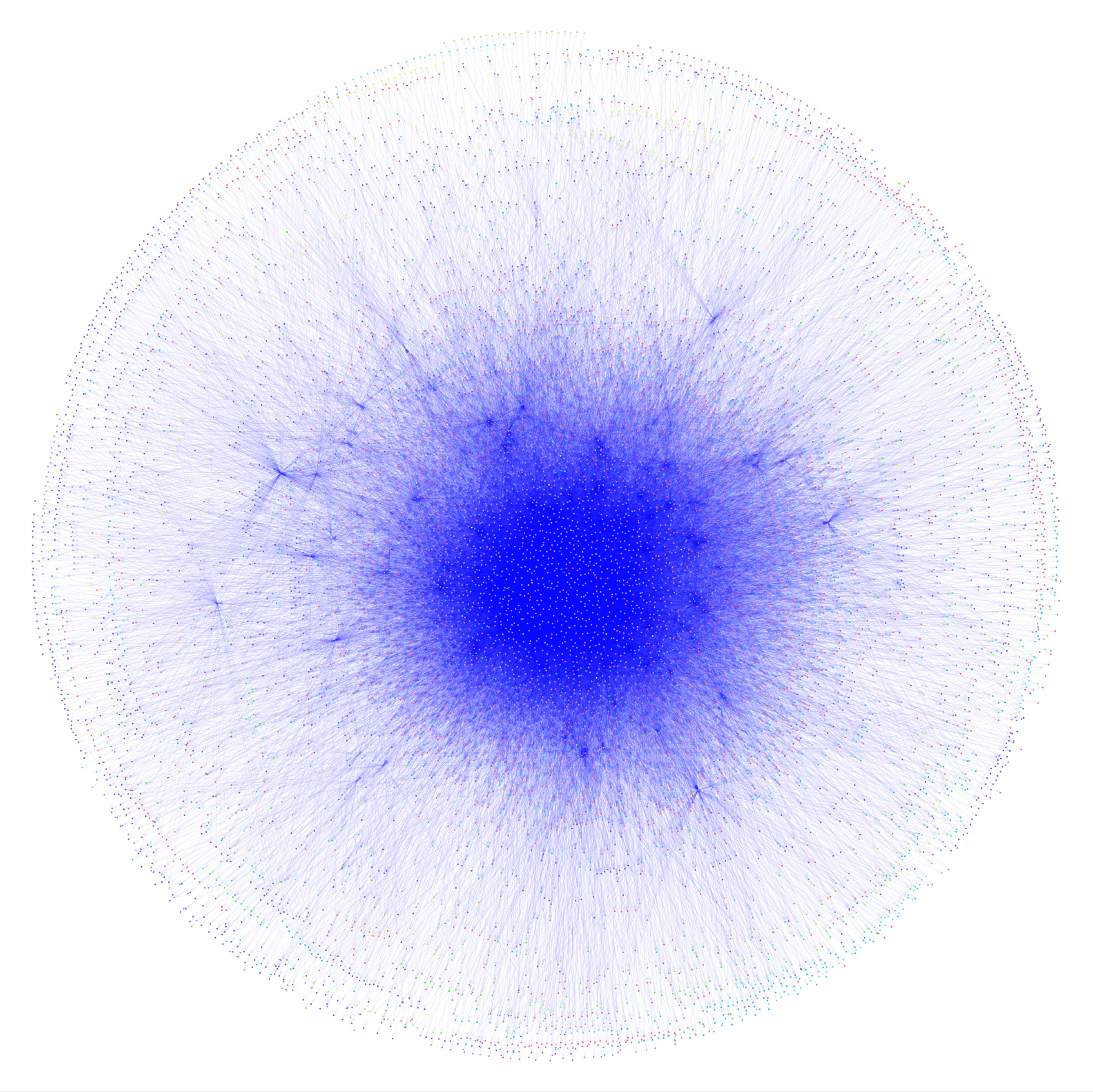}
    \caption{Visualization of EvoBot (human + bot).}
    \label{fig:Evobot}
\end{figure*}

\begin{figure*}[t]
    \centering
    \includegraphics[width=0.95\textwidth,keepaspectratio]{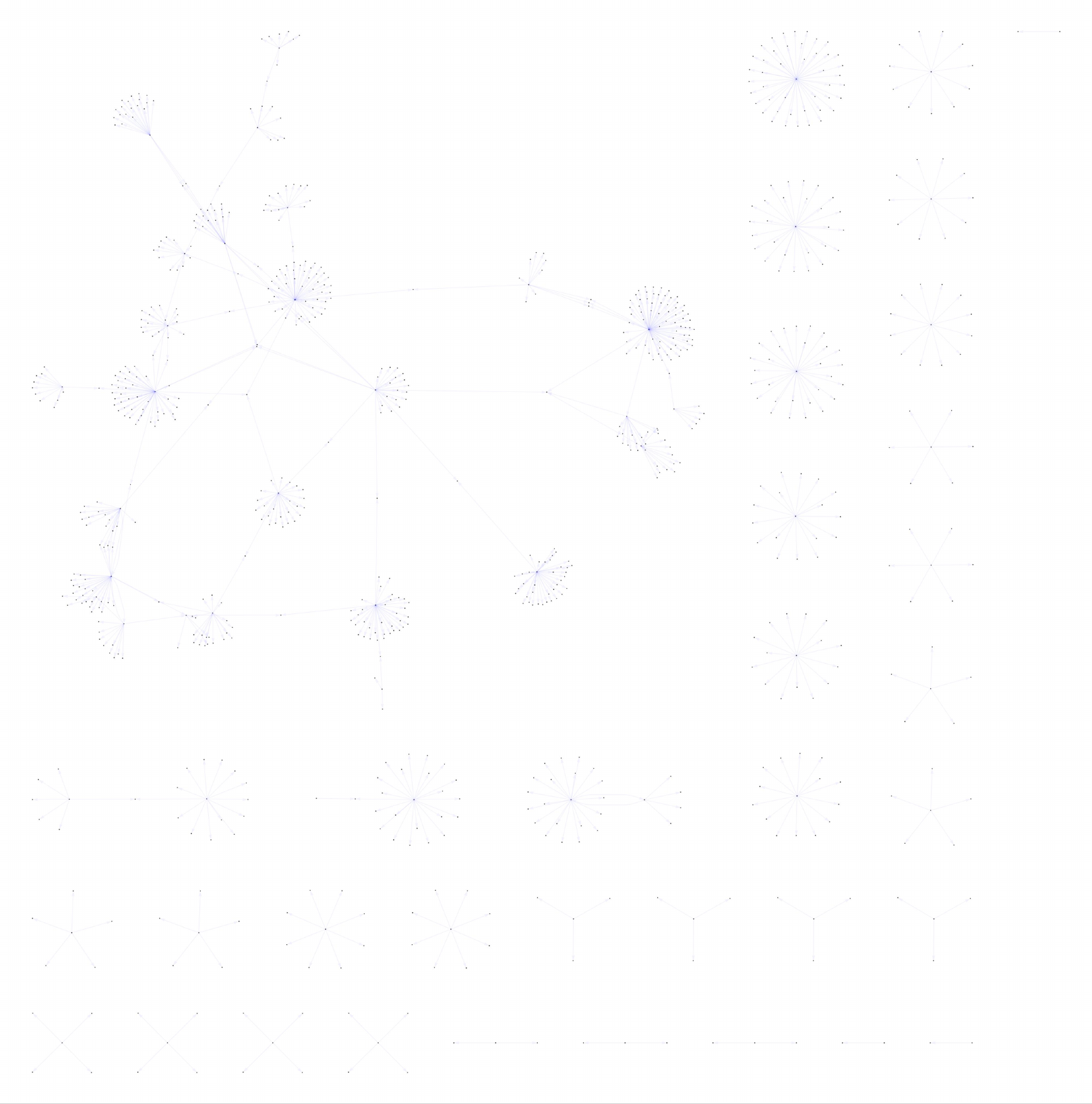}
    \caption{Botnet Visualization of EvoBot.}
    \label{fig:Evobot-botnet}
\end{figure*}


\end{document}